\documentclass{aa} \usepackage[switch]{lineno}

\usepackage{subcaption}
\usepackage{graphicx}
\usepackage{txfonts}
\usepackage{hyperref}

\usepackage{natbib,twoopt}
\bibpunct{(}{)}{;}{a}{}{,}             

\makeatletter
  \newcommandtwoopt{\citeads}[3][][]{\href{http://adsabs.harvard.edu/abs/#3}%
    {\def\hyper@linkstart##1##2{}%
     \let\hyper@linkend\@empty\citealp[#1][#2]{#3}}}
  \newcommandtwoopt{\citepads}[3][][]{\href{http://adsabs.harvard.edu/abs/#3}%
    {\def\hyper@linkstart##1##2{}%
     \let\hyper@linkend\@empty\citep[#1][#2]{#3}}}
  \newcommandtwoopt{\citetads}[3][][]{\href{http://adsabs.harvard.edu/abs/#3}%
    {\def\hyper@linkstart##1##2{}%
     \let\hyper@linkend\@empty\citet[#1][#2]{#3}}}
  \newcommandtwoopt{\citeyearads}[3][][]%
    {\href{http://adsabs.harvard.edu/abs/#3}
    {\def\hyper@linkstart##1##2{}%
     \let\hyper@linkend\@empty\citeyear[#1][#2]{#3}}}
\makeatother

\usepackage{textgreek}
\usepackage{fancyvrb}
\usepackage{amsmath}
\usepackage{upgreek}
\usepackage{wrapfig}
\usepackage{lscape}
\usepackage{epstopdf}
\usepackage{tabularx}
\usepackage{colortbl}
\usepackage[flushleft]{threeparttable}

\newcommand{\gammacyg}{PSR\ J2021+4026}

\newcommand{\gr}{$\gamma$-ray}

\newcommand{\fermi}{\emph{Fermi}-LAT}
\newcommand{\xmm}{XMM-\emph{Newton}}
\newcommand{\degree}{$\hbox{$^\circ$}$}

\definecolor{blue}{RGB}{0, 0, 254}
\definecolor{red}{RGB}{254, 0, 0}
\newcommand\intrev[1]{#1}
\newcommand\wtrev[1]{{#1}}
\newcommand\wtrevtwo[1]{#1}
\newcommand\journalrev[1]{#1}
\newcommand\languagerev[1]{#1}

\newcolumntype{Y}{>{\centering\arraybackslash}X}
\newcolumntype{Z}{>{\hsize=.5\hsize}Y}

\begin{document} 

\title{A phase-resolved \fermi\ analysis of the mode-changing pulsar PSR J2021+4026 shows hints of a multipolar magnetosphere}

\author{
A.~Fiori$^{(1)}$ \and 
M.~Razzano$^{(1)}$ \and 
A.~K.~Harding$^{(2)}$ \and 
M.~Kerr$^{(3)}$ \and 
R.~P.~Mignani$^{(4,5)}$ \and 
P.~M.~Saz~Parkinson$^{(6)}$
}

\institute{
\inst{1}~Universit\`a di Pisa and Istituto Nazionale di Fisica Nucleare, Sezione di Pisa I-56127 Pisa, Italy\\ 
\inst{2}~Los Alamos National Laboratory, Los Alamos, NM 87545, USA\\ 
\inst{3}~Space Science Division, Naval Research Laboratory, Washington, DC 20375-5352, USA\\ 
\inst{4}~INAF-Istituto di Astrofisica Spaziale e Fisica Cosmica Milano, via E. Bassini 15, I-20133 Milano, Italy\\ 
\inst{5}~Janusz Gil Institute of Astronomy, University of Zielona G\'{o}ra, ul Szafrana 2, 65-265, Zielona G\'{o}ra, Poland\\ 
\inst{6}~Santa Cruz Institute for Particle Physics, Department of Physics and Department of Astronomy and Astrophysics, University of California at Santa Cruz, Santa Cruz, CA 95064, USA\\ 
\email{alessio.fiori@pi.infn.it} \\
\email{massimiliano.razzano@unipi.it} \\
}

\date{Submitted December 12, 2023; accepted February 17, 2024}
 
\abstract
    {The radio-quiet \gr\ pulsar PSR J2021+4026 is a peculiar \fermi\ pulsar showing repeated and quasi-periodic mode changes. Its \gr\ flux shows repeated variations between two states at intervals of \wtrevtwo{$\sim 3.5$ years.} These events occur over \languagerev{timescales} $<$100 days and are correlated with sudden changes in the spin-down rate. Multiwavelength observations also revealed an X-ray phase shift relative to the \gr\ profile \wtrevtwo{for} one of the events. PSR J2021+4026 is currently the only known isolated \gr\ pulsar showing significant variability, and thus it has been the object of thorough investigations.}
    {The goal of our work is to study the mode changes of PSR J2021+4026 with improved detail. By accurately characterizing variations in the \gr\ spectrum and pulse profile, we aim to relate the \fermi\ observations to theoretical models. We also aim to interpret the mode changes in terms of variations in the structure of a multipolar dissipative magnetosphere.}
    {We continually monitored the rotational evolution and the \gr\ flux of PSR J2021+4026 using more than 13 years of \fermi\ data with a binned likelihood approach. \intrev{We investigated the features of the phase-resolved spectrum and pulse profile, and from these we inferred the macroscopic conductivity, the electric field parallel to the magnetic field\languagerev{, and} the curvature radiation cutoff energy. These physical quantities are related to the spin-down rate and the \gr\ flux and therefore are relevant to the theoretical interpretation of the mode changes.} We introduced a simple magnetosphere model that combines a dipole field with a strong quadrupole component. We simulated magnetic field configurations to determine the positions of the polar caps for different sets of parameters.}
    {We clearly detect the previous mode changes and \intrev{confirm a more recent} mode change \languagerev{that} occurred around June 2020. We provide a full set of best-fit parameters for the phase-resolved \gr\ spectrum and the pulse profile obtained in five distinct time intervals. We computed the relative variations in the best-fit parameters, finding typical flux changes between $13$\% and $20$\%. Correlations appear between the \gr\ flux and the \intrev{spectral parameters}, as the peak of the spectrum shifts by $\sim$10\% toward lower energies when the flux decreases. 
    The analysis of the pulse profile reveals that the pulsed fraction of the light curve is \intrev{larger} when the flux is low. Finally, the magnetosphere simulations show that some configurations could explain the observed multiwavelength variability. However, self-consistent models are required to reproduce the observed magnitudes of the mode changes.
    }
    {}


\titlerunning{A phase-resolved \fermi\ analysis of PSR J2021+4026}

\maketitle


\section{Introduction}

Since its launch \intrev{aboard} the \wtrevtwo{Fermi Gamma-ray Space Telescope (\emph{Fermi})} mission in 2008, the Large Area Telescope \citep[LAT; ][]{atwood09} \languagerev{has continued} to improve our understanding of pulsar physics. So far, the LAT has detected \intrev{294} \gr\ pulsars now collected in the Third \fermi\ Catalog of \gr\ Pulsars (3PC; \citealt{3pc}), and the number of detections continues to grow\footnote{Public list of LAT-detected \gr\ pulsars: \texttt{https://confluence.slac.stanford.edu/display/GLAMCOG/ Public+List+of+LAT-Detected+Gamma-Ray+Pulsars}}.  \gammacyg\ is a noteworthy radio-quiet \fermi\ pulsar. It is located within the radio shell of the Gamma Cygni supernova remnant (SNR G 78.2+2.1), and therefore it is \intrev{often} referred to as the Gamma Cygni pulsar. However, neither the SNR nor the pulsar are related in any way with the Gamma Cygni supergiant star that is part of the Cygnus constellation. A blind search for periodicity \wtrevtwo{in} the LAT data, using the coordinates of plausible X-ray counterparts of the EGRET source 3EG J2020+4017 \citep{hartman1999, weisskopf2006}\languagerev{, led} to the detection of \gr\ pulsations with a spin period of 265 ms \citep{abdo09}. Subsequent \emph{Chandra} observations led to the identification of the likely X-ray counterpart \citep{weisskopf11}, and X-ray pulsations were later detected with \languagerev{the X-ray Multi-Mirror (XMM-\emph{Newton}) space telescope} \citep{lin13}. \gammacyg\ is a young, energetic, rotation-powered pulsar, with an estimated characteristic age of 77 kyr\footnote{There is a significant discrepancy between the characteristic spin-down age of \gammacyg\ and the estimated age of SNR G 78.2+2.1 of $<$10 kyr \citep{leahy2013}.} and a spin-down luminosity of L$_{\textup{sd}}$ $\sim$ 10$^{35}$ erg s$^{-1}$. Since no radio counterpart has ever been detected, the usual distance based \languagerev{on the} dispersion measure is not possible \citep{2011ray, 2023shaw}. However, a distance of 1.5$\pm$0.4 kpc can be inferred indirectly by its association with SNR G 78.2+2.1 \citep{landecker80}.

\gammacyg\ attracted significant interest in October 2011, when its \intrev{integral \gr\ energy flux above 100 MeV}, $F_\gamma\sim$ 7.9$\times$10$^{-10}$ erg cm$^{-2}$ s$^{-1}$, suddenly decreased by 18$\pm$2\% \intrev{\citep{allafort13}}. The flux decrease was concurrent with an increase in its spin-down rate, $|\dot{f}|\sim\ $7.9$\times$10$^{-13}$ Hz s$^{-1}$, of 5.73$\pm$0.03\%.  No sudden change in the spin frequency was observed. The event occurred over a time scale $<$7 days and the pulsar persisted in this state for more than 3 years. A more gradual recovery phase occurred over $\sim$100 days around December 2014, when the flux and spin-down rate returned to the original state \citep{ng16, zhao17}. A new mode change, similar to the 2011 event, was observed in February 2018 \citep{takata20}. Independent works by \citet{fiori2023} and \citet{prokhorov2023} detected an increase in the \gr\ flux of PSR J2021+4026 around June 2020, \languagerev{which was likely} the recovery stage after the 2018 event. Previous studies (\languagerev{e.g.,} \citealt{allafort13, ng16}) \wtrevtwo{found indications of} variations in the pulse profile and in the hardness of the \gr\ spectrum \wtrevtwo{across mode changes}. Finally, \citet{razzano23} performed a multiwavelength analysis using \fermi\ and \xmm\ data, reporting that the phase lag between the X-ray peak and the brightest \gr\ peak shifted by 0.21$\pm$0.2 rotations across the 2014 mode change. 

Although only four mode changes have been observed so far, the cadence of the events suggests a quasi-periodic variability, with one full cycle lasting about seven years. The change in the \gr\ flux of \gammacyg\ remains unique among \fermi\ pulsars, and the mechanism producing the mode changes is still not fully understood. The variations in the spin-down rate are currently attributed to changes in the structure of the magnetosphere. Crackings and mass displacements on the crust of the neutron star (NS) could change the geometry of the magnetic field near the polar caps \citep{ruderman91}. \citet{ng16} attribute the 2011 event to a change by $\sim$3\degree\ in the magnetic inclination angle. \citet{takata20} suggest that the pinned vorticity model for pulsar glitches may predict the observed \languagerev{timescales} between subsequent events. \citet{razzano23} argue that the presence of a quadrupole component in the magnetic field could explain the phase shift in the X-ray pulsations. Despite this observational and theoretical effort, there remains no conclusive interpretation of the variability of PSR J2021+4026. 

Here we report on updated monitoring observations of \gammacyg, which allowed us to confirm \intrev{the June 2020 mode change reported by \citet{fiori2023}}. We also report the results of an in-depth variability analysis of pulse profiles and phase-resolved spectra. The manuscript is outlined as follows. In Section \ref{sec:2} we provide our data selection criteria. In Section \ref{sec:3} we describe the monitoring of the flux and timing parameters, the spectral analysis\languagerev{, and} the pulse profile analysis. In Section \ref{sec:4} we discuss our observations in terms of a multipolar dissipative magnetosphere. Conclusions follow \intrev{in Section \ref{sec:5}}.


\section{Data reduction}
\label{sec:2}

In this work we used $>$13\,yr of LAT data, collected from August 5, \languagerev{2008,} to October 6, 2021. The data set includes LAT photons of \verb|P8R3_SOURCE_V2| event class \citep{pass8, p8r3}. Data reduction was performed using the official analysis suite, \textit{Fermitools} \intrev{v2.2.0}. We produced a \wtrevtwo{region of interest} (RoI) including all photons within a radius of 10\degree\ of the \wtrevtwo{J2000} \emph{Chandra} position of PSR J2021+4026, RA=20h 21m 32.4s,	Dec=+40\degree\ 26' 40'' \citep{weisskopf11}. \wtrevtwo{We selected data in the energy range from 100 MeV to 300 GeV with zenith angles $z<$ 90\degree\ to avoid contamination from the Earth limb.}  We reduced the data \wtrevtwo{to include only the good time intervals by} applying the filter \verb|DATA_QUAL>0 && LAT_CONFIG==1|. We applied photon barycentering for the timing monitoring \intrev{using the JPLEPH.405 solar system ephemerides \citep{standish1998}}. We then binned the data into 35 logarithmically spaced energy bins (10 bins per decade) and square pixels of size 0\fdg1. \intrev{The LAT has an energy-dependent \wtrevtwo{point spread function} (PSF) affecting the measured directions of incoming photons. Based on the quality of the angular reconstruction, LAT photons are partitioned into four event types (PSF0, PSF1, PSF2, PSF3).} Each event type has a different LAT response; therefore, we produced four sets of binned data.


\section{Variability analysis}
\label{sec:3}

\subsection{Model}
\label{sec:3.1}

We modeled the \gr\ spectrum of \gammacyg\ as a power law with an exponential cutoff,
\begin{linenomath}
\begin{equation}
    \frac{dN}{dE} = 
        K \ \bigg ( \frac{E}{E_0} \bigg )^{-\Gamma + \frac{d}{b}}
        \exp{ \bigg [ \frac{d}{b^2} 
        \bigg (1 - \bigg (\frac{E}{E_0} \bigg )^b \bigg ) \bigg ] }
\label{eq:sed}
\end{equation}
\end{linenomath}
where the reference energy, $E_0$ = 2 GeV, was fixed. This function was introduced in the \fermi\ 12-year Source Catalog (4FGL-DR3; \citealt{4fgl-dr3}) to reduce parameter correlations in the \wtrevtwo{spectral energy distribution} (SED) of LAT pulsars. The normalization factor, $K$, is the photon flux density at the reference energy $E_0$, while the other parameters describe the shape of the spectrum. In particular, \intrev{$\Gamma$} is the power-law index at $E_0$, $d$ is the curvature of the spectrum at $E_0$, and $b$ determines the asymmetry. In the case \wtrevtwo{of} $b < 1$ the cutoff is sub-exponential and the spectrum converges to a power law at low energies. If $b \ll 1$ the spectrum is more symmetric and the convergence to a \intrev{power law} is very slow. In the opposite case, $b > 1$, the cutoff is super-exponential and the convergence to a power law is rapid, therefore the spectrum is very asymmetric.
This set of parameters can be used to infer two physical quantities: the peak energy, \languagerev{that is,} the energy corresponding to the peak of the SED,
\begin{linenomath}
\begin{equation}
     E_p = E_0 \ \bigg [ 1 + \frac{b}{d} (2 - \Gamma) \bigg ] ^\frac{1}{b} \ ,
\label{eq:epeak}
\end{equation}
\end{linenomath}
and the peak curvature, \languagerev{that is,} the second derivative of the SED calculated at the peak energy,
\begin{linenomath}
\begin{equation}
     d_p = d + b \ ( 2 - \Gamma) \ .
\label{eq:dpeak}
\end{equation}
\end{linenomath}
Both are widely discussed in 3PC. \wtrev{We also numerically compute $E_{10}$, defined as the energy at which the SED spectrum is 1/10 of the maximum power at $E_p$ \citep{Kala2022}.} We \languagerev{make} use of these quantities in Section \ref{sec:4} to relate variations in the \wtrev{spectrum} to changes in the properties of the pulsar magnetosphere.

\begin{figure*}[htp]
      \centering
      \includegraphics[width=0.9\hsize]{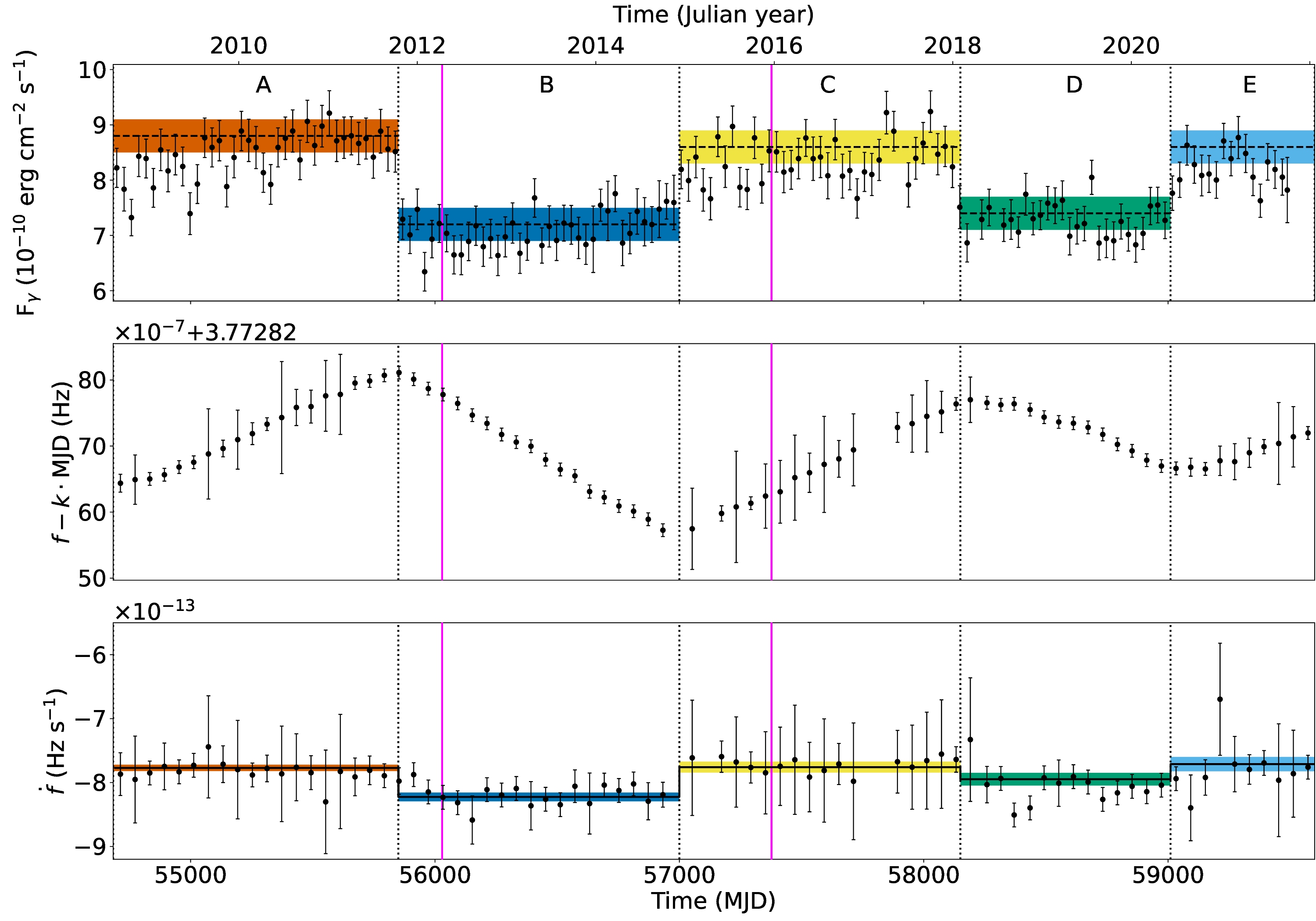}
      \caption{Energy flux and timing parameters of \gammacyg\ in the time range from August 5, \languagerev{2008,} to October 6, 2021. In the top panel, error bars are the result of maximum likelihood fits to 30-day intervals. We show the \intrev{best-fit} values (horizontal dashed lines) and the 3\textsigma\ confidence bands of the flux reported in Table \ref{tab:phase_averaged}, obtained with the phase-averaged spectral analysis (Section \ref{sec:3.3}) in the time intervals A (red), B (blue), C (yellow), D (green)\languagerev{, and }E (cyan). Vertical dashed lines indicate the boundaries of these time intervals. In the mid and bottom panels, error bars are obtained with weighted $H$-tests on 60-day intervals. To enhance the changes in the slope, we report $f - k\ \cdot$ MJD rather than $f$, where $k$ = 6.847$\times 10^{-8}$ Hz day$^{-1}$ is an average spin-down rate obtained from a $\chi^{2}$ fit. The solid line and the colored intervals represent the evolution of the spin-down rate predicted using the parameters of Table \ref{tab:timing_solution}, obtained with the method described in Section \ref{sec:3.4}. We omitted all points with significance $< 5\sigma$. \journalrev{The vertical solid magenta lines at MJD 56028 (April 11, 2012) and MJD 57376 (December 20, 2015) indicate the epochs of the XMM-\textit{Newton} observations analyzed by \citet{razzano23}.}}
      \label{fig:time_series}
\end{figure*}

We built a model of the RoI starting from the 4FGL-DR3 catalog. We included sources within 20\degree\ of \gammacyg\ and templates for the Galactic diffuse emission and the isotropic diffuse emission. In the first instance, we kept all spectral parameters of \gammacyg\ free. We freed the fluxes of other bright pulsars in the RoI (PSR J2021+3651, PSR J2032+4027) and of the supernova remnant G 78.2+2.1. 
We also freed the fluxes of variable sources (\verb|VarIndex > 18.48|) within 7\degree. \intrev{Finally, we freed the normalization and the spectral index of the Galactic diffuse emission, and  we fixed the isotropic diffuse emission.} Applying these criteria, the initial model included 23 free parameters.

\subsection{Flux and timing monitoring}
\label{sec:3.2}

We performed a binned likelihood analysis with summed PSF components using the \emph{pyLikelihood}\footnote{\texttt{https://github.com/fermi-lat/pyLikelihood}} Python suite. The best fit was performed using the NewMinuit optimization algorithm. The accuracy of the fit parameters was enhanced by including likelihood weights, which we computed based on a model for the diffuse background in order to include the \intrev{contributions of} systematic errors. We added two energy bins \intrev{to take energy dispersion into account}; this is expected to increase the accuracy at energies below 1 GeV. As a preliminary step, we ran a best fit to the \wtrevtwo{full} data set \wtrevtwo{(August 2008 -- October 2021)}. This produced the following global parameters: $F_\gamma$ = 8.26$\pm$0.07 $\times$10$^{-10}$ erg cm$^{-2}$ s$^{-1}$, $\Gamma$ = 2.63$\pm$0.01, \intrev{$d$ = 0.81$\pm$0.02}, $b$ = 0.35$\pm$0.04. Using Equations \ref{eq:epeak} and \ref{eq:dpeak}, we infer that $E_p = 800\pm16$ MeV and $d_p = 0.586\pm0.012$. We also computed $E_{10}$, defined as is Section \ref{sec:3.1}, with a numerical approach, and we obtained $E_{10} = 8.9 \pm 0.1$ GeV. 
The energy flux and the exponential parameters $b$ and $d$ are consistent with the values reported in the 4FGL-DR3 catalog \citep{4fgl-dr3}, while the spectral index $\Gamma$ is off by $\sim$$8 \sigma$. This discrepancy is likely due to the longer time span of our data set. In fact, compared to the 4FGL-DR3 analysis, we collected $\sim$1 more year of data, and the increased statistics could have affected the global spectral fit due to the variability of the source. However, the peak energy, $E_{p}$, and curvature, $d_p$, are consistent with the 4FGL-DR3, indicating that the correlation between the parameters of Equation \ref{eq:sed} also plays a role in the discrepancy. We used the set of values reported above as starting parameters for the following analysis.

\begin{table*}[htp]
\centering
\begin{threeparttable}

\caption{\raggedright{Phase-averaged spectral parameters and properties of the pulse profile in different time intervals.}}
\label{tab:phase_averaged}

\begin{tabular}{l>{\columncolor[gray]{0.9}}cc>{\columncolor[gray]{0.9}}cc>{\columncolor[gray]{0.9}}cc}
\hline\hline
	\rowcolor{white}\textbf{Parameter}&\textbf{A}&\textbf{B}&\textbf{C}&\textbf{D}&\textbf{E}\\
	\hline
	MJD start&54683&55850&57000&58150&59010\\
	MJD stop&55850&57000&58150&59010&59493\\
    ISO date start&2008-08-05&2011-10-16&2014-12-09&2018-02-01&2020-06-10 \\
    ISO date stop&2011-10-16&2014-12-09&2018-02-01&2020-06-10&2021-10-06 \\
	Number of days&1167&1150&1150&860&483\\
	$F_\gamma\ $ \tablefootmark{(a)}&$8.8\pm0.1$&$7.2\pm0.1$&$8.6\pm0.1$&$7.4\pm0.1$&$8.6\pm0.1$\\
    $\Gamma$&$2.66\pm0.02$&$2.80\pm0.02$&$2.71\pm0.02$&$2.79\pm0.03$&$2.64\pm0.02$\\
    $d$&$0.81\pm0.03$&$0.90\pm0.05$&$0.82\pm0.04$&$0.85\pm0.04$&$0.76\pm0.04$\\
    $b$&$0.34\pm0.06$&$0.32\pm0.07$&$0.29\pm0.06$&$0.32\pm0.07$&$0.38\pm0.08$\\
    $E_p\ $\tablefootmark{(b)}&$850\pm30$&$770\pm20$&$820\pm20$&$730\pm20$&$790\pm40$\\
$d_p$&$0.59\pm0.02$&$0.64\pm0.02$&$0.61\pm0.02$&$0.60\pm0.02$&$0.52\pm0.03$\\
    $E_{10}\ $\tablefootmark{(c)}&$9.6\pm0.3$&$8.0\pm0.3$&$9.2\pm0.3$&$8.2\pm0.3$&$9.9\pm0.5$\\

    $\delta_{\textup{P1}}\ $\tablefootmark{(d)}&0.142 $\pm$  0.007&0.17 $\pm$  0.01&0.158 $\pm$  0.007&0.118 $\pm$  0.007&0.20 $\pm$  0.01\\
    $\delta_{\textup{BR}}\ $\tablefootmark{(d)}&0.30 $\pm$  0.04&0.16 $\pm$  0.015&0.30 $\pm$  0.05&0.16 $\pm$  0.03&0.30 $\pm$  0.04\\
    $\delta_{\textup{P2}}\ $\tablefootmark{(d)}&0.132 $\pm$  0.003&0.164 $\pm$  0.004&0.132 $\pm$  0.004&0.144 $\pm$  0.002&0.166 $\pm$  0.008\\
    $\Delta_{\textup{P1-BR}}\ $\tablefootmark{(e)}&0.235 $\pm$  0.004&0.271 $\pm$  0.004&0.229 $\pm$  0.005&0.251 $\pm$  0.004&0.27 $\pm$  0.01\\
    $\Delta_{\textup{P1-P2}}\ $\tablefootmark{(e)}&0.532 $\pm$  0.005&0.568 $\pm$  0.004&0.522 $\pm$  0.006&0.544 $\pm$  0.004&0.574 $\pm$  0.011\\
    P1/P2$\ $\tablefootmark{(f)}&0.425 $\pm$  0.019&0.395 $\pm$  0.016&0.46 $\pm$  0.02&0.319 $\pm$  0.015&0.52 $\pm$  0.03\\
    BR/P2$\ $\tablefootmark{(f)}&0.208 $\pm$  0.012&0.19 $\pm$  0.01&0.145 $\pm$  0.011&0.104 $\pm$  0.008&0.27 $\pm$  0.02\\
    const/P2$\ $\tablefootmark{(f)}&0.64 $\pm$  0.02&0.236 $\pm$  0.012&0.85 $\pm$  0.03&0.385 $\pm$  0.017&0.53 $\pm$  0.03\\
	\hline
\end{tabular}

\begin{tablenotes}
\item \tablefoot{Statistical uncertainties only. \intrev{Columns related to high-flux states are highlighted with a \languagerev{gray} background.} \\
\tablefoottext{a}{\gr\ flux $>$ 100 MeV, 10$^{-10}$ erg cm$^{-2}$ s$^{-1}$.}\\
\tablefoottext{b}{MeV.}\\
\tablefoottext{c}{GeV.}\\
\tablefoottext{d}{Peak \intrev{full width at half maximum}.}\\
\tablefoottext{e}{Phase lag between peaks.}\\
\tablefoottext{f}{Ratios of the peak amplitudes or constant-to-peak amplitude ratio.}\\
}
\end{tablenotes}

\end{threeparttable}
\end{table*}

We monitored the \gr\ flux by dividing the full data set into 30-day time windows and analyzing each window independently. At this stage we fixed the spectral parameters of \gammacyg\ to the global ones due to the reduced exposure, while \wtrevtwo{leaving} the flux free to vary. We produced a time series (top panel of Figure \ref{fig:time_series}) spanning the whole 13 years of data. The results reveal a rise in the flux in 2020, which we identified as the recovery after the February 2018 mode change. 
By fitting a piecewise function to the data of Figure \ref{fig:time_series} in the range MJD 58200 - MJD 59493 we found a 13$\pm$2\% flux change between MJD 59000 and MJD 59030. Due to the 30-day bin size, all epochs between these two bin centers are equally valid. Using the $\chi^2$ statistic, we estimated that the piecewise model has a 7.6$\sigma$ significance compared to a constant model. To better characterize the mode change, we also tested a piecewise function with a linear step. The best-fit parameters show that the event is centered on MJD 59010 (June 10, 2020), with a statistical uncertainty of 30 days. The duration of the linear step is about 130 days, while the flux variation is confirmed to be 13$\pm$2\%. The linear step is preferred with 4.4$\sigma$ significance compared to the simple step. We take MJD 59010 as a reference epoch for the following analysis. 

We also monitored the pulsar timing parameters by running a weighted $H_{20}$-test \citep{kerr2011} on 60-day time windows. At this stage, we modeled the phase as \intrev{the fractional part of a second-order} Taylor series,
\begin{linenomath}
\begin{equation}
    \varphi(t) = {\rm frac} \Big [ f(t_0) (t - t_0) + \frac{1}{2} \dot f (t_0) (t - t_0)^2 \Big ] \ ,
    \label{eq:phase}
\end{equation}
\end{linenomath}
where $t_0$ is the center of each time window \intrev{and $\varphi \in [0, 1]$}. Photon probabilities were calculated following the simple weights recipe provided by \citet{bruel2019}. We searched for the optimal weight parameter ($\mu_w$, Equation 11 of \citealt{bruel2019}) with steps of 0.1, and we found that $\mu_w$ = 3.1 maximizes $H_{20}$. We also varied selection radii between 0\fdg8 and 10\degree\ in steps of 0\fdg1 and found that the test significance is maximized by selecting photons within 4\fdg4 of the source.
The tests produced the optimal values of frequency, $f$, and spin-down rate, $\dot f$, as a function of time (middle and bottom panels of Figure \ref{fig:time_series}). Uncertainties were estimated by bootstrapping samples of the same size for each time interval, repeating the scan and taking the standard deviation of the optimal parameters. We observe that error bars are generally larger when the flux is higher: assuming that 100\% of the \gr\ flux is pulsed, this suggests a change in the significance of the pulsation across the mode changes. 
Although there are clear changes in the slope of $f$, 
a direct fit to the points of the time series does not provide a good precision on the $\dot f$ changes, due to the large uncertainties in the high-flux intervals. Instead, a global timing solution (Section \ref{sec:3.4}) \languagerev{provides} a better resolution on $\dot f$.

\subsection{Phase-averaged spectral analysis}
\label{sec:3.3}

\begin{figure*}[htp]
    \centering
    \begin{subfigure}{0.8\hsize}
        \includegraphics[width=0.49\hsize]{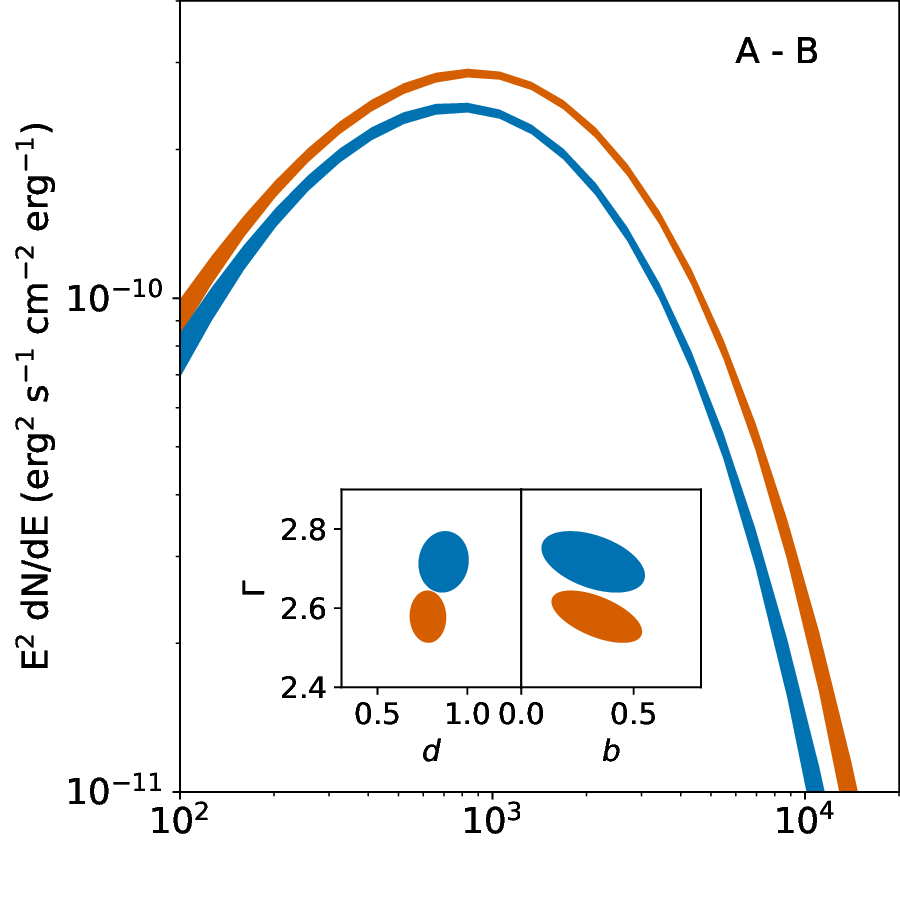}
        \includegraphics[width=0.49\hsize]{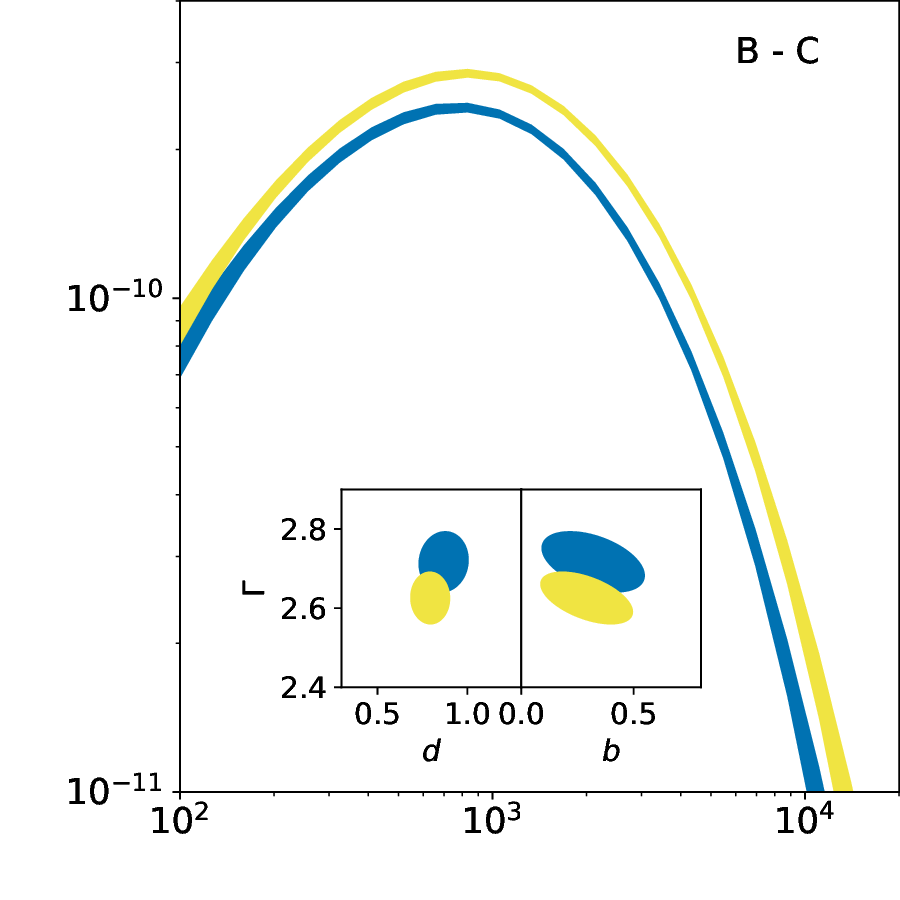}
    \end{subfigure}
    \begin{subfigure}{0.8\hsize}
    \includegraphics[width=0.49\hsize]{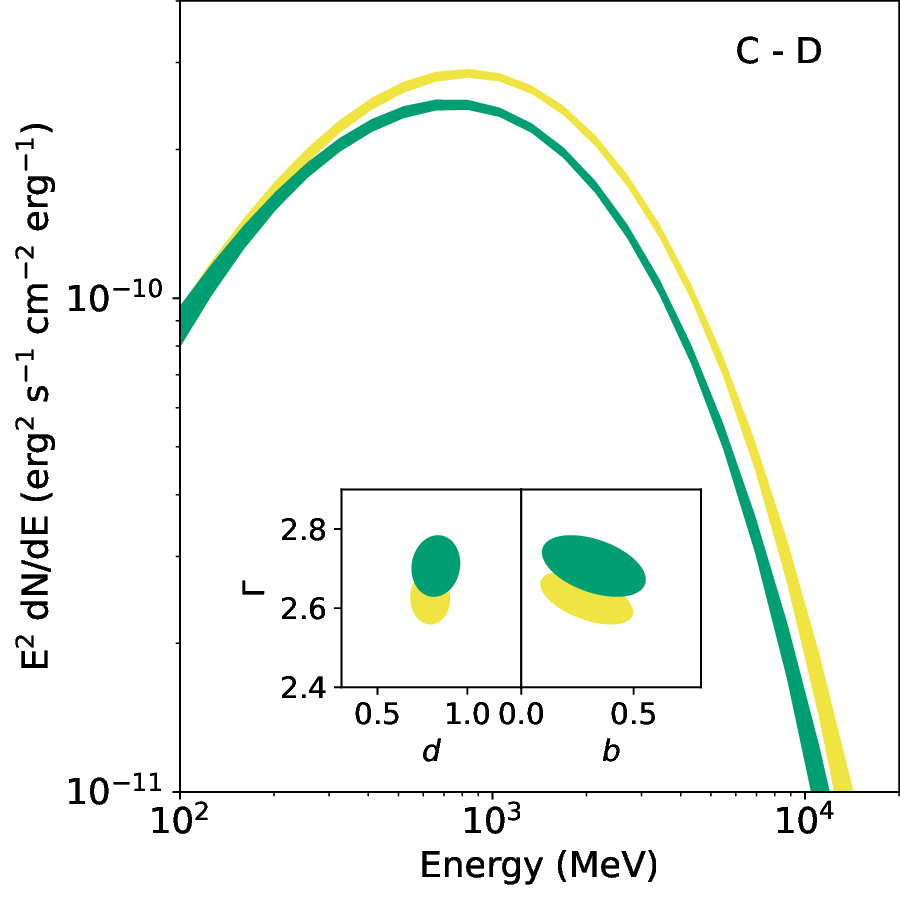}
    \includegraphics[width=0.49\hsize]{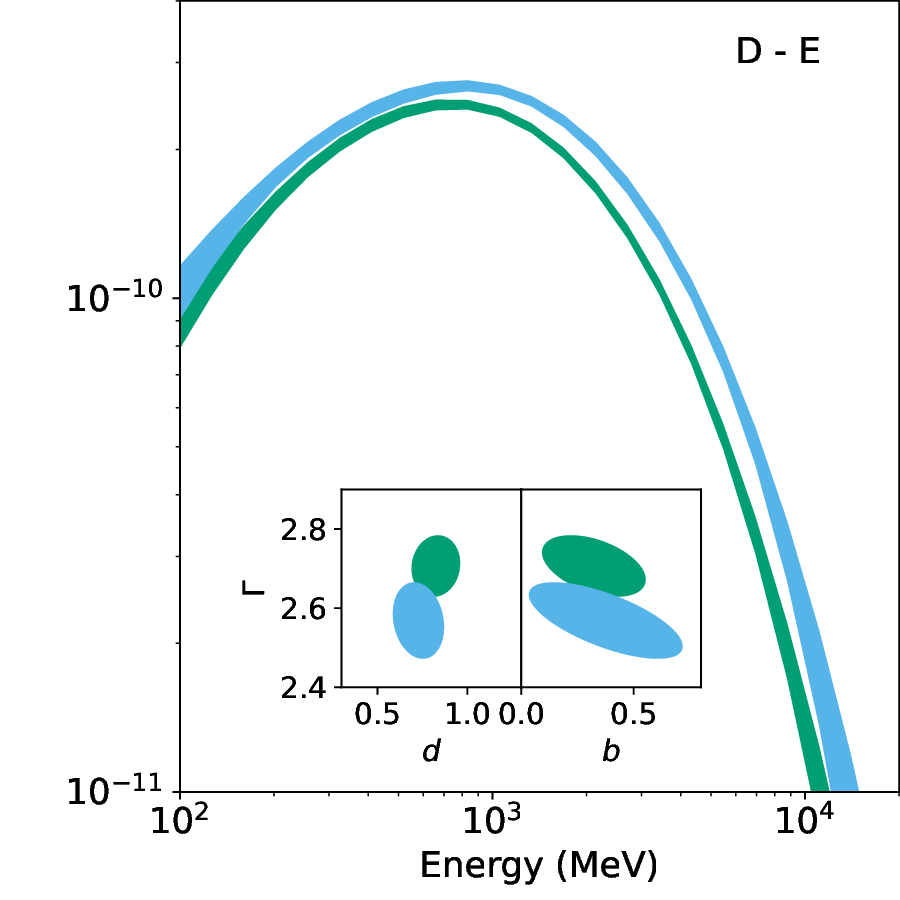}
    \end{subfigure}
    \caption{Fitted SEDs of PSR J2021+4026 in intervals A (red), B (blue), C (yellow), D (green)\languagerev{, and }E (cyan) at the four mode changes. The bands represent the 3\textsigma\ confidence intervals from a multivariate Gaussian distribution. The inset panels show the 3\textsigma\ confidence ellipses around the optimal values of the spectral parameters.}
    \label{fig:spectrum}
\end{figure*}

To study the changes in the spectral properties, we divided our full data set into five distinct time intervals, labeled A--E. These intervals were defined according to the previously observed events \citep{allafort13, ng16, takata20} \wtrevtwo{as well as the most recent event} reported here. The boundaries of the intervals are given in Table \ref{tab:phase_averaged}.
We performed the same binned likelihood analysis presented in Section \ref{sec:3.2} on the five intervals. We kept all the spectral parameters of PSR J2021+4026 free. The results are provided in Table \ref{tab:phase_averaged}. The best-fit values of the energy flux and its 3$\sigma$ confidence bands are shown in the top panel of Figure \ref{fig:time_series} for comparison with the monitoring results.

We used the results of the fit to infer the relative flux changes at the four events. We define (\languagerev{e.g.,} for the first mode change) $\Delta F_\gamma / F_\gamma$(A--B) = [$F_\gamma$(B) -- $F_\gamma$(A)] / $F_\gamma$(A), \languagerev{that is,} the ratio between the change in $F_\gamma$ and the value of $F_\gamma$ in the previous state. The four relative flux changes are the \intrev{following:} $\Delta F_\gamma / F_\gamma$(A--B) = --18.1$\pm$1.5\%, $\Delta F_\gamma / F_\gamma$(B--C) = 20$\pm$2\%, $\Delta F_\gamma / F_\gamma$(C--D) = --13.6$\pm$1.6\% \intrev{and} $\Delta F_\gamma / F_\gamma$(D--E) = 16$\pm$2\%. At each transition from low to high \gr\ flux, $F_\gamma$ always returns to the previous high-flux value within 2$\sigma$, \languagerev{that is,} $F_\gamma$(A) $\sim$ $F_\gamma$(C) $\sim$ $F_\gamma$(E). The two low-flux values are also comparable, \languagerev{that is,} $F_\gamma$(B) $\sim$ $F_\gamma$(D). In all intervals, the test statistic indicates that our model has a significance > 5$\sigma$ with respect to a model with constant flux. Indications of spectral softening occur when the flux drops. We observe that the relative increase of $\Gamma$ is 5.3$\pm$1.1\% at A--B and 2.9$\pm$1.3\% at C--D, respectively. Comparable but opposite variations can be observed for the 2014 and 2020 events, in particular --3$\pm$1\% at B--C and --5.4$\pm$1.2\% ad D--E. The other parameters, $d$ and $b$, do not appear to change significantly. 
For this reason, we also tested a model with $b$ fixed to the value obtained from the global fit, finding that a variable $b$ is preferred with respect to a model with constant $b$ in all the intervals at the $>3\sigma$ level. 
For each time interval, we sampled the multivariate Gaussian distribution defined by the best-fit parameters and the covariance matrix produced by the fit. \wtrev{We computed samples for $E_p$, $d_p$\languagerev{, and }$E_{10}$, and we} estimated their average values and standard deviations. The results reveal that when the flux is low the SED is peaked at lower energies. The relative change in $E_p$ is about --10\% at both the A--B and C--D mode changes,\wtrev{ while $E_{10}$ varies by --16\% at A--B and by --10\% at C--D}. On the other hand, the spectral curvature $d_p$ does not follow the same trend and is consistent with a constant value. Figure \ref{fig:spectrum} shows the pre-change and post-change best-fit SED at the four events. 


\subsection{Timing analysis and pulse profile}
\label{sec:3.4}

 \begin{table*}[htp]
\centering
\begin{threeparttable}

\caption{\raggedright{Best-fit parameters of the timing model.}}
\label{tab:timing_solution}

\begin{tabular}{lccccc}
\hline\hline
\textbf{Parameter}& &\textbf{A--B}&\textbf{B--C}&\textbf{C--D}&\textbf{D--E}\\
\hline
MJD Validity Range&54690.20 $-$ 59600.47&&&&\\
Reference MJD Epoch&57083.77&&&&\\
$f\ $\tablefootmark{(a)}&3.7689233(5)&&&&\\
$\dot f\ $\tablefootmark{(b)}&$-$7.77(5)&&&&\\
Glitch MJD Epoch&&55850.00&57000.00&58150.00&59010.00\\
$\Delta \dot f\ $\tablefootmark{(c)}&&$-$4.5(5)&4.7(5)&$-$1.9(5)&2.4(5)\\
$\Delta \dot f / \dot f\ $\tablefootmark{(d)}&&0.058(6)&$-$0.057(6)&0.024(6)&$-$0.030(6)\\
$\log{\mathcal{L}}$&6409.2&&&&\\
\hline
\end{tabular}

\begin{tablenotes}
\item \tablefoot{We report statistical errors on the last digit in parentheses.\\
\tablefoottext{a}{Frequency at the reference epoch, Hz.}\\
\tablefoottext{b}{Frequency derivative at the reference epoch, 10$^{-13}$ Hz s$^{-1}$.}\\
\tablefoottext{c}{\intrev{Change} in $\dot f$ at the glitch epoch, 10$^{-14}$ Hz s$^{-1}$.}\\
\tablefoottext{d}{Inferred relative change in $\dot f$ computed at the glitch epoch.}\\
}
\end{tablenotes}

\end{threeparttable}
\end{table*}

 We produced a timing model using an unbinned maximum likelihood approach \citep{ajello2022}. We simultaneously \intrev{fit} the timing model with a model for a stationary noise process whose power spectral density is assumed to follow a \intrev{power law}. The timing model includes terms in the Taylor series up to the second derivative of the frequency, $\ddot f$, although the best-fit value of $\ddot f$ is consistent with zero. \intrev{To fit the mode changes,} we added four glitches, modeled as permanent changes in the frequency derivative, $\Delta \dot f$.
 Due to the significant level of the timing noise in this pulsar \citep{razzano23}, we were not able to fit the epochs of the glitches. Therefore, we manually set them to the MJD of the flux mode changes and kept them fixed. Changes in the absolute phase and in the frequency at the glitches are consistent with zero, therefore we did not include them in the model. The noise model is represented by a truncated Fourier series with 50 terms. The best-fit approach is the same \intrev{as used in} \citet{razzano23}, but here we extended the validity range of the timing solution to January 2022. The best-fit parameters are reported in Table \ref{tab:timing_solution}. 
 The bottom panel of Figure \ref{fig:time_series} shows the best-fit values of $\dot f$ and its 3$\sigma$ confidence bands, consistently with how we represented the \gr\ flux in the top panel. For each glitch we also inferred $\Delta \dot f / \dot f$. For consistency with our definition of $\Delta F_\gamma / F_\gamma$, we defined this quantity as the ratio between the glitch $\Delta \dot f$ and the value of $\dot f$ in the previous state. We estimated the uncertainties by sampling points from the timing parameter space, computing the ratio for each point, then taking the standard deviation of this distribution.
 
In order to characterize the pulse profile, we calculated photon weights based on the full-mission spectral model of \gammacyg\ (Section \ref{sec:3.2}). We performed an unbinned likelihood fit using the template fitting module included in the PINT Python package \citep{pint}. We modeled the profile with the sum of $m$ phase-wrapped Gaussian components and a constant unpulsed term,
\begin{linenomath}
\begin{equation}
\frac{dN}{d\varphi} \propto a_0 + 
\sum_{i=1}^{m} \frac{a_i}{\sqrt{2 \pi \sigma_i^2}} 
\exp \bigg [ {\frac{1}{2} \frac{(\varphi - \mu_i)^2}{\sigma_i^2}} \bigg ] \ ,
\label{eq:profile}
\end{equation}
\end{linenomath}
where we \languagerev{imposed} the normalization condition
\begin{linenomath}
\begin{equation}
\sum_{i=0}^{m} a_i = 1\ .
\label{eq:norm}
\end{equation}
\end{linenomath}
We tested different choices of $m$ and found that $m$=3 is the simplest model that well describes the pulse profile. A model with $m$=4 has a test significance < 3$\sigma$ when compared with $m$=3. We performed two preliminary fits by freeing first only the pulsed-to-constant component ratio, then only the peak locations. This produced a set of starting values for the fit. We then fixed the central peak location and ran the main fit algorithm with all other parameters free. We underline that a unique fit with 10 free parameters could not be performed due to strong parameter correlations. This procedure was applied to each of the five time intervals. We also tested a model with four Gaussian peaks and found that the significance of the extra peak was below 3 sigma. Therefore, we concluded that the three-Gaussian model properly describes the pulse shape.

\begin{figure}[htp]
  \centering
  \includegraphics[width=\hsize]{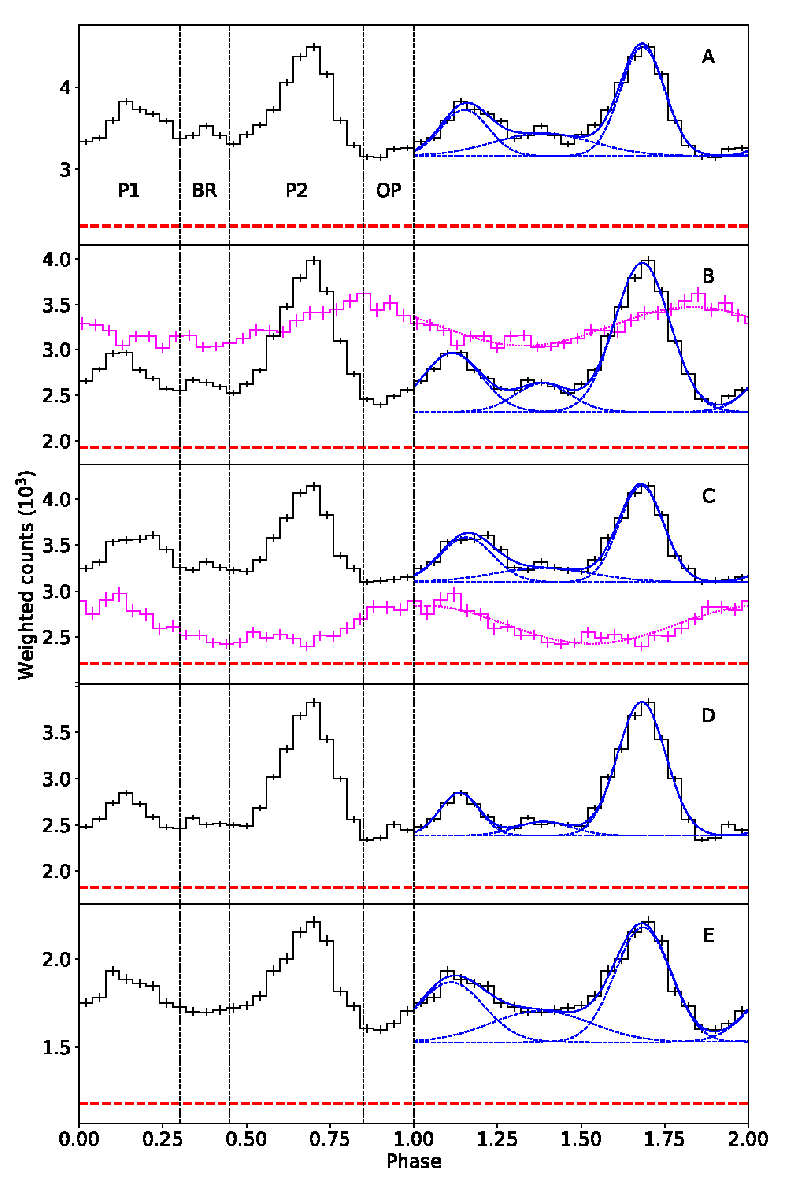}
  \caption{Pulse profile of PSR J2021+4026 in different time intervals. \journalrev{Black solid lines represent the weighted histograms of \gr\ counts produced with phase bins of size 0.04.} Only statistical uncertainties are reported. Blue solid lines show the \journalrev{\gr} best-fit functions. Blue dashed lines represent single Gaussian components. The \journalrev{\gr} background is estimated from the photon probabilities and is reported with red dashed lines. Vertical dashed lines represent the boundaries of the phase intervals defined in Section \ref{sec:3.5}. \journalrev{For the sake of completeness, we include the X-ray histograms (magenta solid lines) and best-fit functions (magenta dotted lines) as reported by \citet{razzano23}. The scales and offsets of the X-ray curves are arbitrary.}}
  \label{fig:pulse_prof}
\end{figure}

The phase-folded pulse profile (Figure \ref{fig:pulse_prof}) reveals two peaks (P1, P2), the second of which is the highest. \journalrev{The two main peaks are separated by $\sim$0.5 in phase, and they are linked by a wide emission that is similar in shape to the third Vela peak \citep{vela}. For consistency with previous works \citep{allafort13, ng16} we refer to this third peak as bridge emission (BR).} The peaks sit upon a bright constant component. 
The fraction of flux produced by the peaks is $\sim$0.3, \languagerev{which means} it is smaller than the contribution from the unpulsed component. The off-pulse (OP) region of the light curve, located between P2 and P1, is very narrow and covers $<$15\% of the rotational period. \journalrev{The unusually broad peaks have been previously explained by the emission geometry required to fit the \gr\ pulsations. For example, \citet{pierbattista2015} tested different emission models and concluded that the observed light curve is likely the result of either a small magnetic inclination angle with a large viewing angle, or vice versa. This is also consistent with the pulsar being radio-quiet.}

The results \journalrev{of our likelihood fit} (Table \ref{tab:phase_averaged}) show that the mode changes affect the overall shape of the pulses. For instance, all amplitude ratios with respect to P2 are lower whenever the \gr\ flux is low, \languagerev{that is,} in intervals B and D. The most evident example of this effect is the change in the \intrev{const-to-P2 ratio}, that varies by up to 70\% in 2014. \intrev{Because the amplitude of the constant relative to the main peak decreases when the pulsar switches to a low-flux state, }the pulsation significance is higher in intervals B and D, as hinted by the timing monitoring (Section \ref{sec:3.2}). \intrev{The separation between P1 and P2 undergoes small variations but no evident correlation with the flux. However, the largest change in the P1--P2 phase lag is $\sim$8\% and occurs again in 2014.} There are also indications of variability in the widths of the peaks. In particular, when the flux is low P2 has a larger width, while the bridge emission appears to be narrower. It is difficult to draw conclusions about variability of the bridge component, as its parameters have large uncertainties and are affected by correlations with other parameters.


\subsection{Phase-resolved spectral analysis}
\label{sec:3.5}

We searched for more detailed changes by performing a phase-resolved spectral analysis. We subdivided each time interval in four phase ($\varphi$) regions: 0--0.3 (P1); 0.3--0.45 (BR); 0.45--0.85 (P2); 0.85--1 (OP). These regions were arbitrarily defined in order to collect at least 95\% of the flux from peaks P1 and P2 in all time intervals. Limited phase ranges were reserved to the bridge emission and the off-peak component, as choosing wider boundaries would have increased the contributions from the main peaks. We repeated the spectral analysis outlined in Section \ref{sec:3.3} on each phase region independently. At this stage, due to the reduced exposure, we were not able to fit the $b$ parameter. Therefore, we kept it fixed to the value from the full-mission phase-averaged fit, $b$ = 0.35. The results are reported in Table \ref{tab:phase_resolved}. \par

\begin{table*}[htp]
\centering
\begin{threeparttable}

\caption{\raggedright{Phase-resolved spectral parameters in different time intervals and in different phase regions.}}
\label{tab:phase_resolved}

\begin{tabular}{l>{\columncolor[gray]{0.9}}cc>{\columncolor[gray]{0.9}}cc>{\columncolor[gray]{0.9}}cc}
\hline\hline
\rowcolor{white}\textbf{Parameter}&\textbf{A}&\textbf{B}&\textbf{C}&\textbf{D}&\textbf{E}\\
\hline
MJD start&54683&55850&57000&58150&59010\\
MJD stop&55850&57000&58150&59010&59493\\
ISO date start&2008-08-05&2011-10-16&2014-12-09&2018-02-01&2020-06-10 \\
ISO date stop&2011-10-16&2014-12-09&2018-02-01&2020-06-10&2021-10-06 \\
Number of days&1167&1150&1150&860&483\\

\hline
\rowcolor{white}&&\multicolumn{2}{c}{\textbf{P1 ($\boldsymbol{\Delta \varphi}$ = 0.3)}}&&\\
\hline

$F_\gamma\ $\tablefootmark{(a)}&$2.65\pm0.05$&$1.90\pm0.04$&$2.64\pm0.05$&$1.88\pm0.04$&$2.59\pm0.06$\\
$\Gamma$&$2.52\pm0.02$&$2.69\pm0.04$&$2.58\pm0.03$&$2.72\pm0.05$&$2.52\pm0.03$\\
$d$&$0.72\pm0.03$&$0.79\pm0.05$&$0.76\pm0.03$&$0.80\pm0.05$&$0.70\pm0.04$\\
$E_p\ $ \tablefootmark{(b)}&$860\pm30$&$710\pm30$&$830\pm30$&$680\pm30$&$850\pm40$\\
$d_p$&$0.53\pm0.03$&$0.55\pm0.04$&$0.56\pm0.03$&$0.54\pm0.04$&$0.51\pm0.03$\\
$E_{10}\ $\tablefootmark{(c)}&$10.6\pm0.5$&$8.4\pm0.5$&$9.7\pm0.4$&$8.2\pm0.6$&$10.8\pm0.7$\\

\hline
\rowcolor{white}&&\multicolumn{2}{c}{\textbf{BR ($\boldsymbol{\Delta \varphi}$ = 0.15)}}&&\\
\hline

$F_\gamma\ $\tablefootmark{(a)}&$1.14\pm0.03$&$0.76\pm0.03$&$1.12\pm0.03$&$0.80\pm0.03$&$1.06\pm0.04$\\
$\Gamma$&$2.49\pm0.04$&$2.83\pm0.07$&$2.57\pm0.04$&$2.76\pm0.06$&$2.54\pm0.05$\\
$d$&$0.71\pm0.05$&$0.97\pm0.09$&$0.80\pm0.05$&$0.83\pm0.07$&$0.68\pm0.07$\\
$E_p\ $ \tablefootmark{(b)}&$910\pm40$&$720\pm40$&$870\pm40$&$660\pm40$&$780\pm60$\\
$d_p$&$0.54\pm0.04$&$0.67\pm0.06$&$0.60\pm0.04$&$0.56\pm0.06$&$0.48\pm0.05$\\
$E_{10}\ $\tablefootmark{(c)}&$11.1\pm0.8$&$7.0\pm0.6$&$9.6\pm0.6$&$7.7\pm0.7$&$10.7\pm1.2$\\

\hline
\rowcolor{white}&&\multicolumn{2}{c}{\textbf{P2 ($\boldsymbol{\Delta \varphi}$ = 0.4)}}&&\\
\hline

$F_\gamma\ $\tablefootmark{(a)}&$4.05\pm0.06$&$3.86\pm0.06$&$3.85\pm0.06$&$3.98\pm0.06$&$3.95\pm0.07$\\
$\Gamma$&$2.59\pm0.02$&$2.67\pm0.02$&$2.57\pm0.02$&$2.67\pm0.02$&$2.58\pm0.03$\\
$d$&$0.84\pm0.03$&$0.90\pm0.03$&$0.82\pm0.03$&$0.87\pm0.03$&$0.74\pm0.04$\\
$E_p\ $ \tablefootmark{(b)}&$900\pm20$&$850\pm20$&$900\pm20$&$820\pm20$&$810\pm30$\\
$d_p$&$0.64\pm0.02$&$0.67\pm0.03$&$0.62\pm0.02$&$0.63\pm0.02$&$0.54\pm0.03$\\
$E_{10}\ $\tablefootmark{(c)}&$9.2\pm0.3$&$8.2\pm0.3$&$9.4\pm0.3$&$8.4\pm0.3$&$9.8\pm0.5$\\

\hline
\rowcolor{white}&&\multicolumn{2}{c}{\textbf{OP ($\boldsymbol{\Delta \varphi}$ = 0.15)}}&&\\
\hline

$F_\gamma\ $ \tablefootmark{(a)}&$0.85\pm0.03$&$0.59\pm0.02$&$0.89\pm0.14$&$0.68\pm0.03$&$0.89\pm0.04$\\
$\Gamma$&$2.77\pm0.07$&$2.95\pm0.08$&$2.98\pm0.06$&$2.76\pm0.09$&$2.65\pm0.08$\\
$d$&$0.85\pm0.08$&$0.92\pm0.09$&$1.03\pm0.07$&$0.8\pm0.1$&$0.73\pm0.09$\\
$E_p\ $ \tablefootmark{(b)}&$670\pm40$&$560\pm40$&$630\pm120$&$630\pm50$&$680\pm60$\\
$d_p$&$0.57\pm0.06$&$0.59\pm0.07$&$0.68\pm0.09$&$0.54\pm0.07$&$0.50\pm0.07$\\
$E_{10}\ $\tablefootmark{(c)}&$7.6\pm0.7$&$6.2\pm0.5$&$5.9\pm0.4$&$8\pm1$&$9.2\pm1.3$\\

\hline
\end{tabular}

\begin{tablenotes}
\item \tablefoot{Statistical uncertainties only. \intrev{Columns related to high-flux states are highlighted with a \languagerev{gray} background.}\\
\tablefoottext{a}{\gr\ flux $>$ 100 MeV, 10$^{-10}$ erg cm$^{-2}$ s$^{-1}$.}\\
\tablefoottext{b}{MeV.}\\
\tablefoottext{c}{GeV.}
}
\end{tablenotes}
\end{threeparttable}
\end{table*}

\begin{figure}[tp]
  \centering
  \includegraphics[width=\hsize]{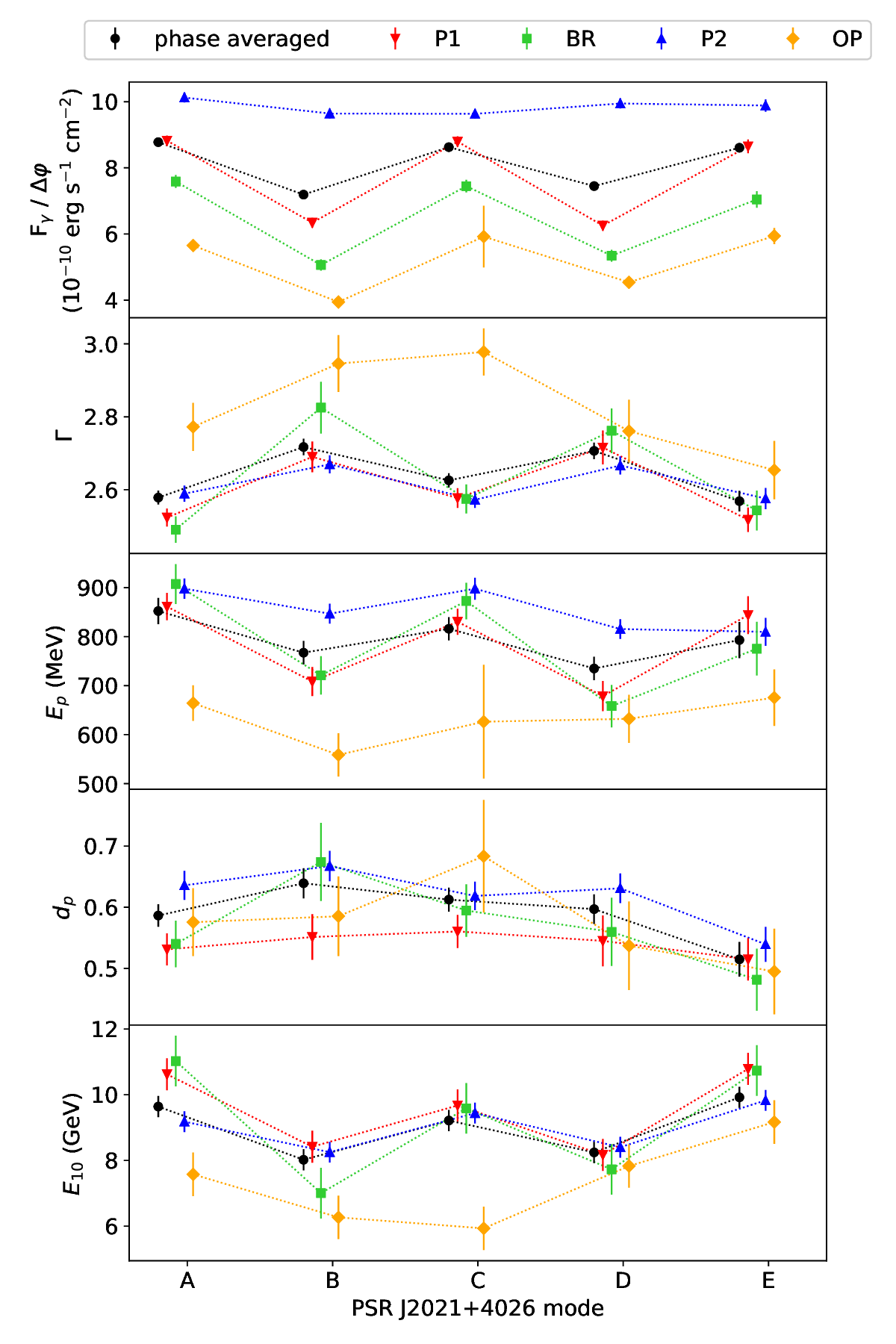}
  \caption{Measured physical parameters obtained from the phase-averaged and phase-resolved spectral analysis in different time intervals. Values and uncertainties are reported in Tables \ref{tab:phase_averaged} and \ref{tab:phase_resolved}. Error bars are grouped according to the time intervals and connected by dotted lines of consistent color. The horizontal axis is not to scale.}
  \label{fig:parameters}
\end{figure}

Figure \ref{fig:parameters} presents the flux and spectral index as obtained from the phase-resolved fit in all time intervals. The largest contribution to the overall flux decrease comes from the P1 region ($\sim$46\% in 2011, $\sim$63\% in 2018). On the other hand, the flux of P2 is stable and does not contribute significantly to the overall change. The flux evolution follows the same trend observed in the phase-averaged analysis in all phase regions except P2. The spectral index is not significantly phase-dependent, with \wtrevtwo{the exceptions of intervals B and C} where the OP spectrum appears softer. Because the uncertainties are large, this apparent change may be a statistical fluctuation. We also report the inferred values of $E_p$, $d_p$ \wtrev{and $E_{10}$} in Figure \ref{fig:parameters}. We observe that the trend of the $E_p$ \wtrev{and $E_{10}$} variations is consistent with the phase-averaged analysis. The P1 and BR spectra show the largest relative $E_p$ jumps, while in OP the value of the parameter is systematically lower. \wtrev{We observe consistent variations of $E_{10}$ in P1, BR\languagerev{, and }P2, while the spectrum of OP has a lower $E_{10}$ at all phases.} Finally, $d_p$ appears to be phase-independent and remains time-independent at all phases.


\section{Discussion}
\label{sec:4}

We have presented an analysis of \wtrev{\gammacyg, the only known variable \gr\ pulsar,} using \fermi\ data spanning more than 13 years. In this time period, we observed two full cycles. Each cycle consists of a first mode change, where the \gr\ flux decreased and the frequency derivative increased, followed by a second mode change  after $\sim$3 years \intrev{nearly returning to }the previous state. The mode changes do not conform to the typical glitch behavior seen in many rotation-powered pulsars, which usually \intrev{involves} a \intrev{positive} change in frequency. \intrev{Instead, \gammacyg\ shows no sudden spin up at the mode changes. Moreover, the typical jump in $\dot f$ is smaller compared to the values measured for \gammacyg, and no \gr\ flux change is observed in correlation with the usual glitches.}  \citet{takata20} discuss several models for the \gammacyg\ mode changes, including a change in the magnetic inclination angle, a change in the magnetospheric current, a change in surface magnetic field structure, precession of the NS\languagerev{, and }a magnetar twisted-field model, as well as a standard pulsar glitch model.  They favor a change in surface magnetic field structure\languagerev{, which} could be due to stresses on the crust that cause cracking and movement of the surface field lines.  Figure 6 of \citet{takata20} shows that the $\Delta \dot f$ and $\dot f$ observed for \gammacyg\ lie on the $\Delta \dot f$ vs. $\dot f$ correlation seen for both pulsar glitches and pulsars showing correlated $\dot f$ and radio pulse profile changes \citep{lyne10}, suggesting that the \gammacyg\ mode changes could be related to the \citet{lyne10} mode changes and could also involve crustal shifts.

\journalrev{The X-ray pulsations of \gammacyg\ exhibit one broad peak (panels B and C of Figure \ref{fig:pulse_prof}). With a multiwavelength timing analysis, \citet{razzano23} revealed that the position of the X-ray peak relative to highest \gr\ peak (P2) shifted by $0.21\pm0.02$ across the 2014 mode change. They also reported that the X-ray spectrum in the 0.3--12 keV energy range is well described by a two-component model defined as a blackbody plus a power law. In particular, the blackbody component has a flux of $(11.0\pm2.4)\times10^{-14}$ erg cm$^{-2}$ s$^{-1}$, while the power-law component has a flux of $(3.2\pm0.7)\times10^{-14}$ erg cm$^{-2}$ s$^{-1}$. This implies that the thermal emission dominates the X-ray flux, and $\sim$80\% of the X rays must be emitted at the NS surface. \citet{razzano23} outlined a possible model for the change in $\dot f$ and \gr\ flux at the B--C mode change, as a change in the magnetospheric conductivity (or equivalently, the pair multiplicity). To explain the results of their analysis, as well as a possible change in pair multiplicity, they speculated that} a changing quadrupole field component near the NS surface could cause the current sheet of the dipole field, which is dominant far from the NS, to connect to a different pole of the quadrupole.  Different conditions for pair production at the different poles of the quadrupole could provide higher or lower conductivity and consequently change $\dot f$ and the \gr\ flux.  Here, we develop this idea more quantitatively.


To physically connect the changes in $\dot f$ to the conductivity, we invoke results of global pulsar magnetosphere models.  Dissipative global models assume a macroscopic conductivity $\sigma$ that relates the \intrev{current density $\boldsymbol{J}$} to the electric field parallel to the magnetic field, $\boldsymbol{E}_\parallel$ \citep{Kala2012},
\begin{linenomath}
\begin{equation} \label{eq:J}
\boldsymbol{J} = c\rho\,\frac{\boldsymbol{E} \times \boldsymbol{B}}{B^2}\,+\sigma \boldsymbol{E}_\parallel
\end{equation}
\end{linenomath}
where $\rho$ is the charge density and $\boldsymbol{E}$ and $\boldsymbol{B}$ are the total electric and magnetic fields.  The first term in Equation \ref{eq:J} is the drift velocity contribution to the current.  \citet{li12} give an expression (in their Equation 9) for the pulsar spin-down luminosity $L $ in terms of the $\sigma$ of their dissipative global models,
\begin{linenomath}
\begin{equation}
\frac{L}{L_0} = 0.3+0.3\log\left(\frac{\sigma}{\Omega}\right)^2+1.2\sin^2\alpha
\end{equation}
\end{linenomath}
for relatively high $\sigma$ near to the force-free (FF) case, where $L_0$ is $3/2$ times the spin-down power of a vacuum rotator, $\Omega$ is the pulsar rotation rate\languagerev{, and }$\alpha$ is the magnetic inclination angle.
\intrev{We assume that $\Omega$ is constant since there is no observed change in $f$ at the mode changes. Because the variations in the shape and position of the peaks are small, we also assume that $\alpha$ does not change.  Writing $L = -4\pi^2 I\,f\,\dot f$, we can deduce that $\Delta \dot f / \dot f = 0.6\ \Omega\ (L_0 / L)\ \Delta \sigma/\sigma$. In the near-FF condition we approximate $L/L_0 \sim 1 + \sin^2 \alpha$ \citep{li12}, and setting $\alpha=45^\circ$, which would reproduce the $\gamma$-ray profile \wtrevtwo{\citep{Kala2014}}, we get $\Delta \dot f / \dot f = 2.5 f\ \Delta \sigma/\sigma$.}
Using the observed frequency $f$ and the fractional changes $\Delta \dot f/ \dot f$ listed in Table \ref{tab:timing_solution} we get the values of $\Delta \sigma/\sigma$ reported in Table \ref{tab:inferred}.

\begin{table}[tp]
\centering
\begin{threeparttable}

\caption{\raggedright{Measured and inferred parameters for the dissipative magnetosphere model.}}
\label{tab:inferred}

\begin{tabularx}{\linewidth}{XYY}
\hline\hline
\textbf{Parameter}&\textbf{A--B}&\textbf{C--D}\\
\hline
$\Delta \dot f / \dot f$ \ \tablefootmark{(meas)}&0.058$\pm$0.006&0.024$\pm$0.06\\
$\Delta \sigma / \sigma$&0.006&0.003\\
$\Delta E_\parallel / E_\parallel$&--0.006&--0.003\\
$\Delta E_{\rm CR} / E_{\rm CR}$&--0.005&--0.002\\
$\Delta \Gamma / \Gamma$ \ \tablefootmark{(meas)}&0.053$\pm$0.011&0.029$\pm$0.013\\
$\Delta E_{\rm p} / E_{\rm p}$&--0.24&--0.13\\
$\Delta E_{\rm p} / E_{\rm p}$ \ \tablefootmark{(meas)}&--0.09$\pm$0.03&--0.11$\pm$0.03\\
$\Delta F_\gamma / F_\gamma$&0.17&0.09\\
$\Delta F_\gamma / F_\gamma$ \ \tablefootmark{(meas)}&--0.181$\pm$0.0136&--0.136$\pm$0.016\\
$\Delta E_{10} / E_{10}$\tablefootmark{(meas)}&--0.16$\pm$0.04&--0.10$\pm$0.04\\
$\Delta L_\gamma / L_\gamma$&--0.18&--0.12\\
\hline
\end{tabularx}

\begin{tablenotes}
\item \tablefoot{Parameters are defined in Section \ref{sec:4}. We omit statistical uncertainties for parameters that are inferred from the model.\\
\tablefoottext{meas}{Measured values as reported in Tables \ref{tab:phase_averaged} and \ref{tab:timing_solution}.}\\
}
\end{tablenotes}

\end{threeparttable}
\end{table}

The macroscopic $\sigma$ in the dissipative models physically reflects the density of pair plasma in the magnetosphere that is able to screen the $\boldsymbol{E}_\parallel$ and reach closer to \intrev{an FF} state where $\sigma \rightarrow \infty$.  An increase in $\sigma$ in the A--B and C--D mode transitions then implies a magnetosphere slightly closer to FF.  In a near-FF magnetosphere, a state that it is believed all \fermi\ pulsars must approach \citep{Kala2014}, the current will adjust to the FF current $\boldsymbol{J}_{\rm FF}$ with any small change in $\sigma$.  It has been shown that the global current controls the polar cap pair creation rather than the pair creation controlling the global current \citep{Timokhin2010}.  Therefore, Equation \ref{eq:J} implies that $\Delta E_\parallel / E_\parallel = -\Delta \sigma/\sigma$ if $\boldsymbol{J} = \boldsymbol{J}_{\rm FF}$ is constant. Consequently, increases in conductivity will screen more of the $\boldsymbol{E}_\parallel$, assuming that $\rho$ and the field strengths do not change. \intrev{The observed changes in the P1-to-P2 peak amplitude ratio suggest a change also in the spatial distribution of $\boldsymbol{E}_\parallel$ affecting the emissivity asymmetrically, or variations in the local magnetic field structure.}

A decrease in $\boldsymbol{E}_\parallel$ will decrease the particle acceleration and thus, potentially the $\gamma$-ray luminosity.  Both global dissipative models and more recent kinetic plasma (particle-in-cell, PIC) simulations of global pulsar magnetospheres \citep{Kala2018,Philippov2018} show that in the near-FF state, most particle acceleration takes place near the current sheet beyond the light cylinder.  The current sheet is therefore the site of the $\gamma$-ray emission, \intrev{and the variations in the width of the peaks observed across the \gammacyg\ mode changes may be due to changes in the size of the acceleration regions.}  However, the $\gamma$-ray emission mechanism is currently under debate, with some favoring curvature radiation \citep{Kala2018} and others favoring synchrotron radiation \citep{Philippov2018} for the GeV emission.  This disagreement results from the different interpretations of the PIC models\languagerev{, which} cannot simulate realistically high  values of pulsar surface magnetic field strengths. \wtrev{It has been shown that all \fermi\ pulsars lie on the same fundamental plane if they are emitting CR near the light cylinder \citep{Kala2019,Kala2022}. Therefore, }for the estimates of this paper, we assume curvature radiation (CR) for the GeV emission of \gammacyg.  Moreover, we assume that the CR emitting particles have reached the radiation-reaction limit, where the energy gain from $\boldsymbol{E}_\parallel$ acceleration balances the energy loss to CR.  In CR reaction limit, which we assume holds in this case, the CR cutoff energy in the spectrum $E_{\rm CR} \propto E^{3/4}_\parallel$ \citep{Kala2019}. Therefore, from the dependency derived above, $\Delta E_\parallel / E_\parallel = -\Delta \sigma/\sigma$, we have $\Delta E_{\rm CR}/E_{\rm CR} = (3/4)\, \Delta E_\parallel / E_\parallel$.  

The cutoff energy can be made explicit in the spectral photon distribution with a different choice of parameters (Equation 14 in 3PC). If we assume that the cutoff energy of the parametric function $dN / dE$ is approximately the CR cutoff energy, then from Equations 17 and 21 of 3PC we get a relation between $E_{\rm p}$, $E_{\rm CR}$ and the spectral parameters. Since there are no indications of changes in $b$ and $d$, we assume that $E_{\rm p}$ is only a function of $E_{\rm CR}$ and $\Gamma$. With a linear approximation we get
\begin{linenomath}
\begin{equation}
    \frac{\Delta E_{\rm p}}{E_{\rm p}} = 
    \frac{\Delta E_{\rm CR}}{E_{\rm CR}} - 
    \frac{1}{b} \ \Bigg [ \frac{\Gamma}{2 - \Gamma + d / b} \Bigg ] 
    \frac{\Delta \Gamma}{\Gamma} \ 
\end{equation}
\end{linenomath}
and we can infer the predicted changes $ \Delta E_{\rm p} / E_{\rm p}$ from the $\Delta E_{\rm CR}/E_{\rm CR}$ computed above and from the measured $\Delta \Gamma / \Gamma$. We can compare our predictions (Table \ref{tab:inferred}) with the observed changes in phase-averaged $E_{\rm p}$ from the values listed in Table \ref{tab:phase_averaged} and shown in Figure \ref{fig:parameters}. The predicted changes are in the same direction as the observed variations. The predicted value for the $(B$--$C)$ change agrees with that observed within the errors, while the prediction for the $(A$-$B)$ change is off by $\sim$2$\sigma$.

Next, we define $F_\gamma = \int E (dN/dE) dE$, and we note that it can be related to $E_{\rm p}$ through Equations \ref{eq:sed} and \ref{eq:epeak}. We can therefore derive the change in $F_\gamma$ as a function of $\Delta E_{\rm p}/E_{\rm p}$,
\begin{linenomath}
\begin{equation}
    \frac{\Delta F_\gamma}{F_\gamma} = 
    \Lambda \, \Bigg ( \frac{E_{\rm p}}{E_0} \Bigg )^b \, \frac{\Delta E_{\rm p}}{E_{\rm p}} \ ,
    \label{eq:df/f}
\end{equation}
\end{linenomath}
where
\begin{linenomath}
\begin{equation}
    \Lambda = \frac{d}{b} \log \Bigg [ \Bigg ( \frac{E_{\rm p}}{E_0} \Bigg )^b - \frac{b^2}{d} \Bigg ] \, + \,
    \frac{b}{2}\, \Bigg [ \Bigg ( \frac{E_{\rm p}}{E_0} \Bigg )^b - \frac{b^2}{d} \Bigg ]^{-1} \ .
\end{equation}
\end{linenomath}
In Equation \ref{eq:df/f}, if we take $\Delta E_{\rm p}/E_{\rm p}$ from our predicted $E_{\rm CR}$ changes given above, and $d$, $b$\languagerev{, and }$E_{\rm p}$ from the global values reported in Section \ref{sec:3.2}, we can estimate the fractional $\gamma$-ray flux changes $\Delta F_\gamma/F_\gamma$. The predictions show that the effect of a negative $\Delta E_{\rm p}/E_{\rm p}$ produces variations that are smaller in amplitude and of opposite sign compared to the observed $\Delta F_\gamma/F_\gamma$. We also added a term $\propto \Delta d / d$ to Equation \ref{eq:df/f} and we proved that the results do not change significantly. 
\journalrev{This implies that we cannot quantitatively reproduce the measured flux changes if we only assume fluctuations in the conductivity. Therefore, we must allow for variations in other quantities.} 

We can state that in a near-FF magnetosphere, where $\boldsymbol J$ stays constant to the FF value, the measured change in $F_\gamma$ is likely the result of variations in the effective magnetic field (Equation \ref{eq:J}). We recall the theoretical fundamental plane relation in the CR-regime, $L_\gamma \propto E_{\rm CR}^{4/3}\ B_*^{1/6}\ L^{5/12}$, where $L_\gamma$ is the \gr\ luminosity, $E_{\rm CR}$ is the spectral cutoff energy, $B_*$ is the surface magnetic field\languagerev{, and }$L = -4\pi^2 I\,f\,\dot f$  \citep{Kala2019}. As pointed out by \citet{Kala2022}, the high-energy part of the \gr\ spectrum is well probed by the $E_{10}$ characteristic energy, so that $\Delta E_{10} / E_{10} \sim \Delta E_{\rm CR} / E_{\rm CR}$. Moreover, assuming dipole radiation in vacuum, we can approximate $\Delta B_* / B_* \sim 0.5\ \Delta \dot f / \dot f$, \languagerev{corresponding to} a small change in $B_*$ by $\sim$2.9\% in A--B, $\sim$1.2\% in C--D. We can thus write
\begin{linenomath}
\begin{equation}
    \frac{\Delta L_\gamma}{L_\gamma} \, = \,
    \frac{4}{3} \Bigg ( \frac{\Delta E_{10}}{ E_{10}} \Bigg )\, + \,
    \frac{1}{2} \Bigg ( \frac{\Delta \dot f}{\dot f} \Bigg ) \ ,
\end{equation}
\end{linenomath}
and using the measured quantities reported in Table \ref{tab:phase_averaged} we obtain the $\Delta L_\gamma / L_\gamma$ of Table \ref{tab:inferred}. The inferred changes are remarkably close to the measured $\Delta F_\gamma / F_\gamma$, and this implies that a field reconfiguration must occur in order for the pulsar to remain on the fundamental plane across the mode changes.

\begin{figure*}[t]
    \centering
    \begin{subfigure}{0.8\hsize}
        \fbox{
            \begin{subfigure}{0.48\hsize}
                \includegraphics[width=\hsize]{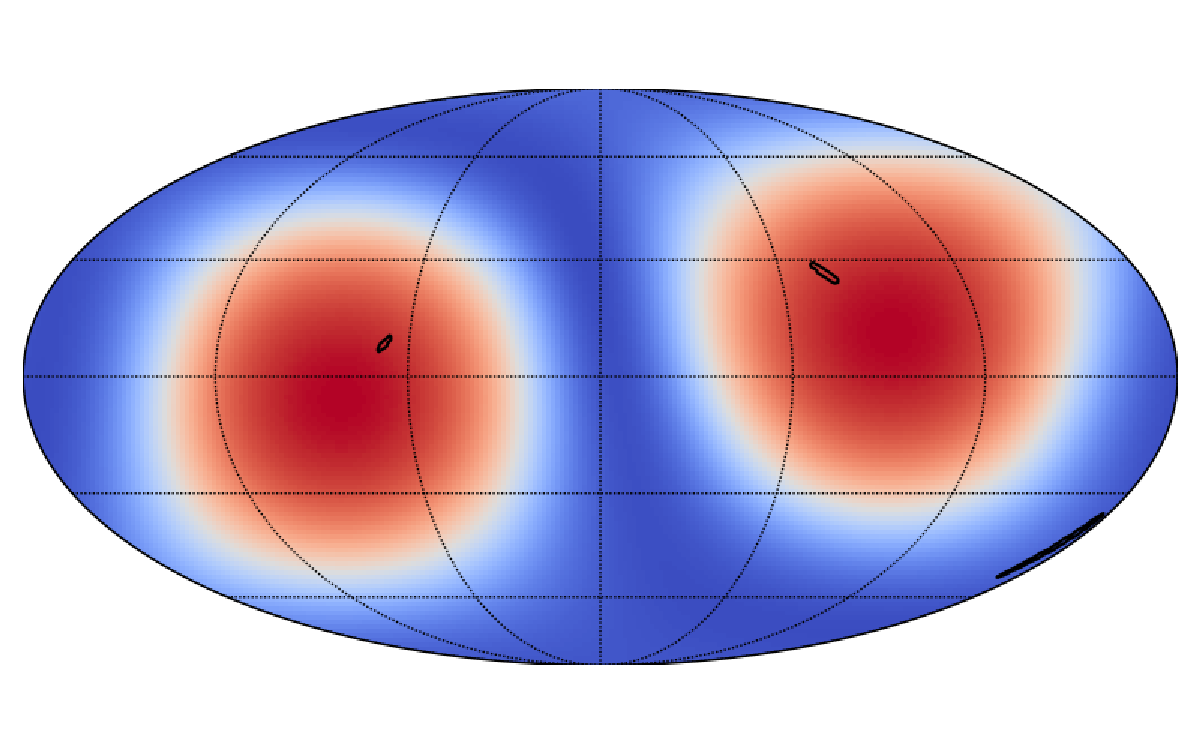}\\
                \includegraphics[width=\hsize]{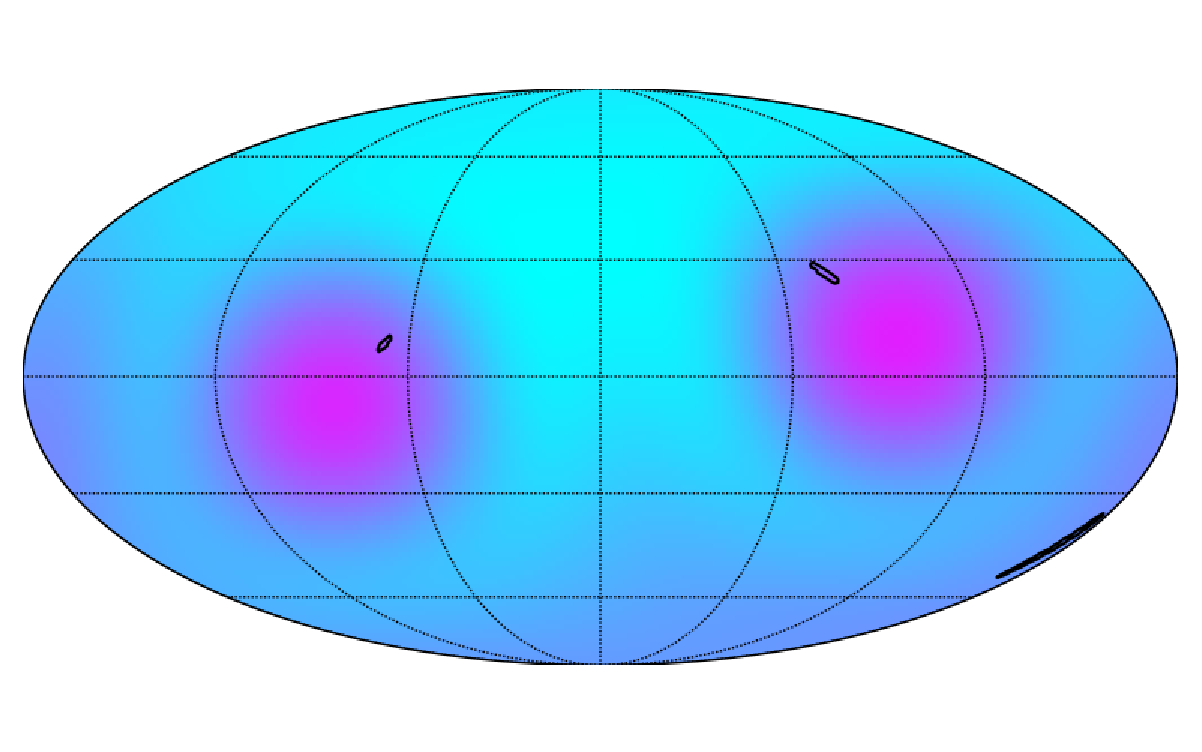}\\
                \includegraphics[width=\hsize]{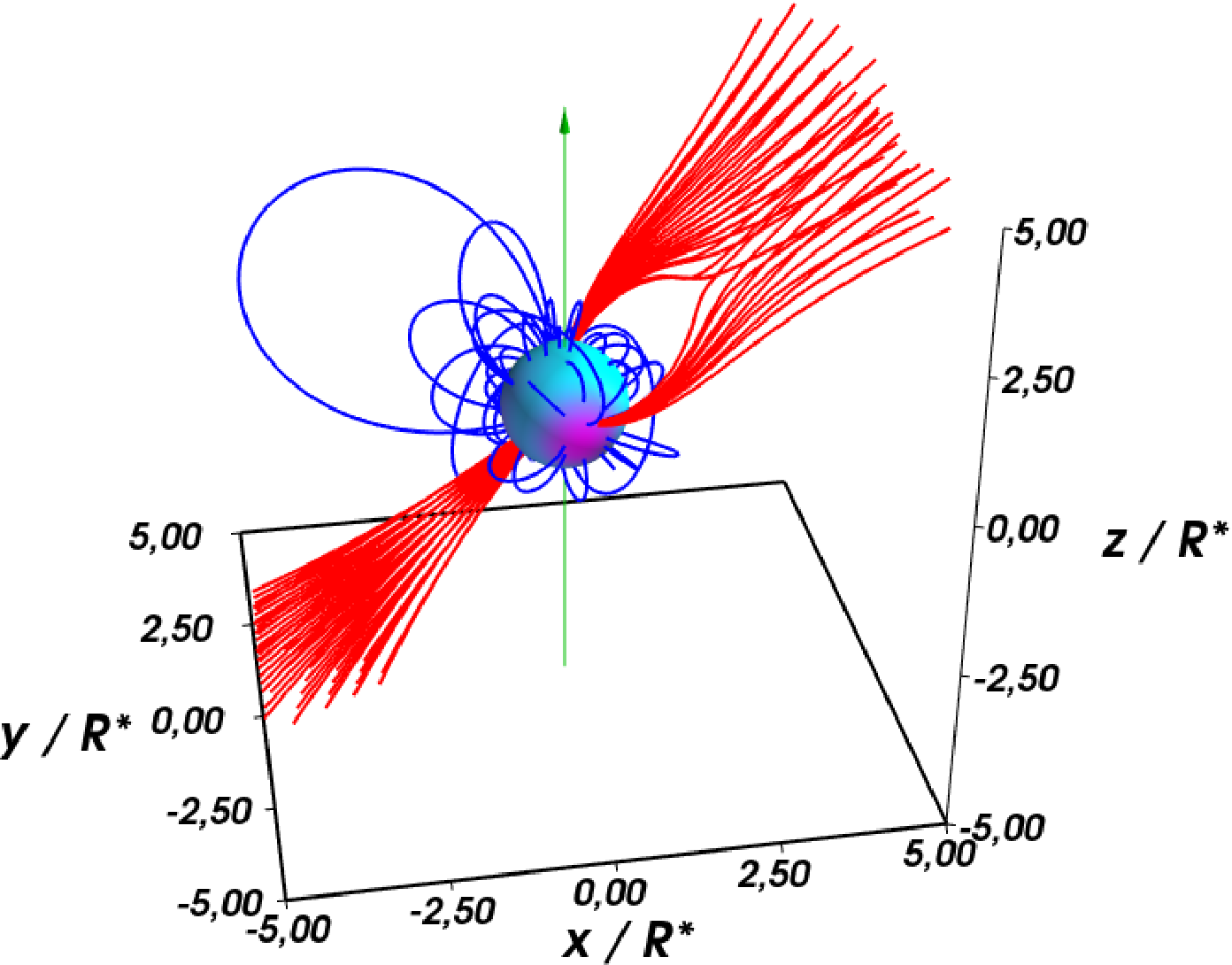}
                \hspace{20pt}
            \end{subfigure}
        }
        \fbox{
            \begin{subfigure}{0.48\hsize}
                \includegraphics[width=\hsize]{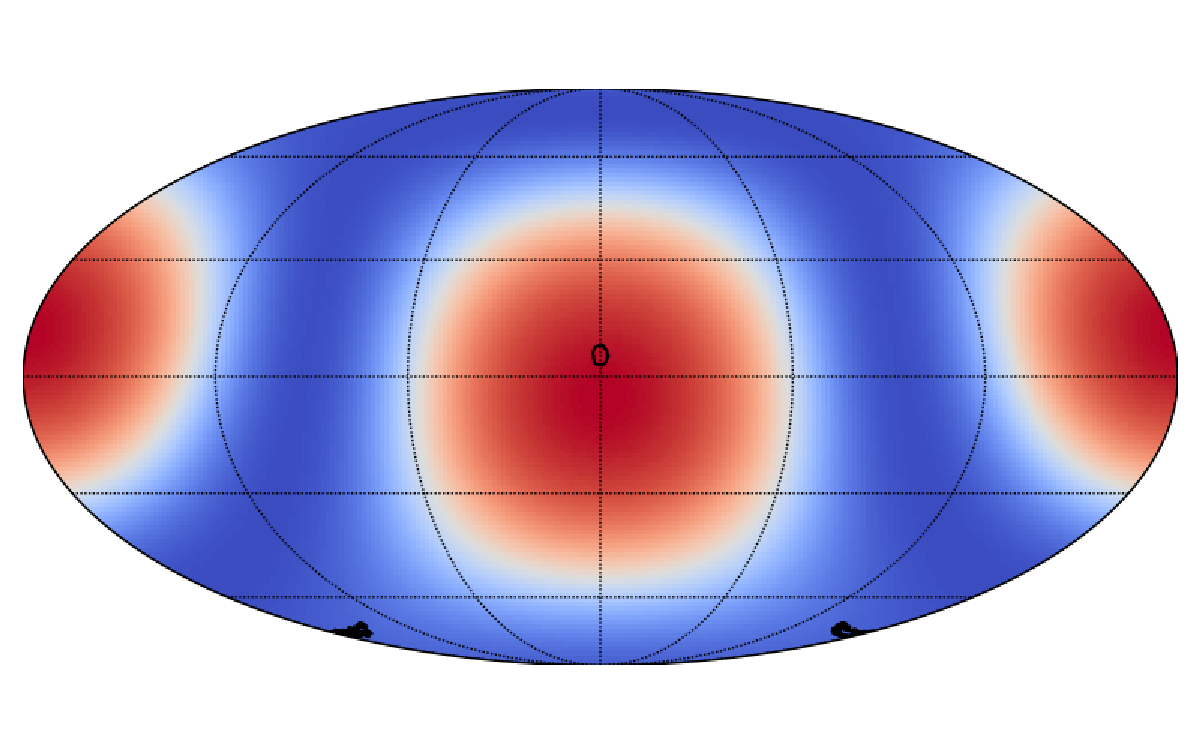}\\
                \includegraphics[width=\hsize]{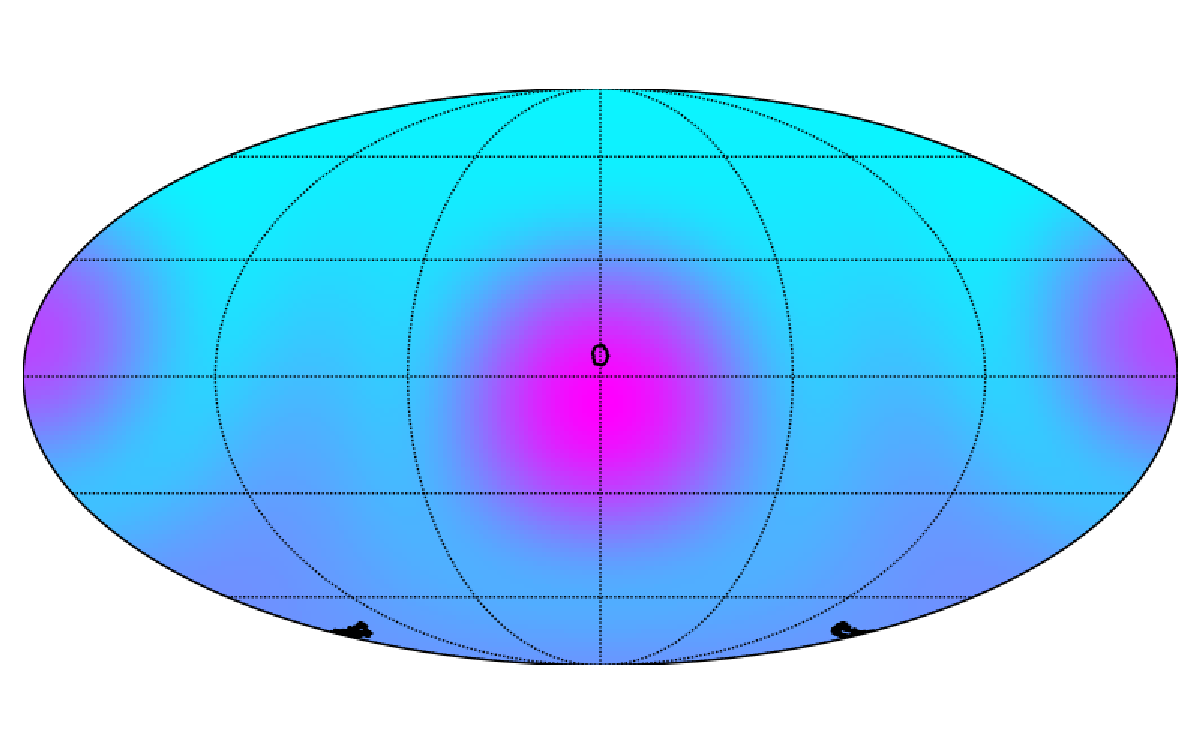}\\
                \includegraphics[width=\hsize]{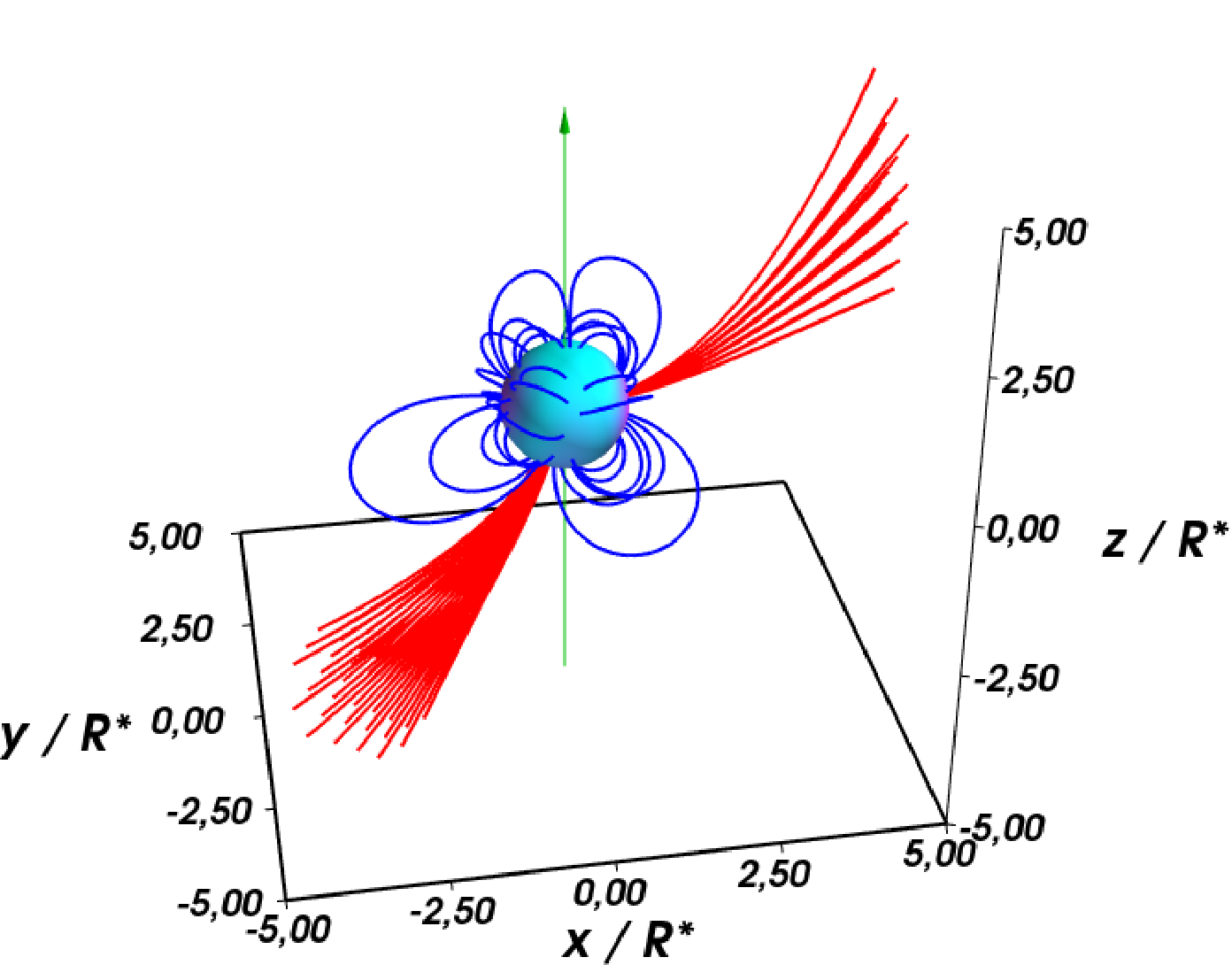}
                \hspace{20pt}
            \end{subfigure}
        }
        \begin{subfigure}{\hsize}
            \includegraphics[width=0.49\hsize]{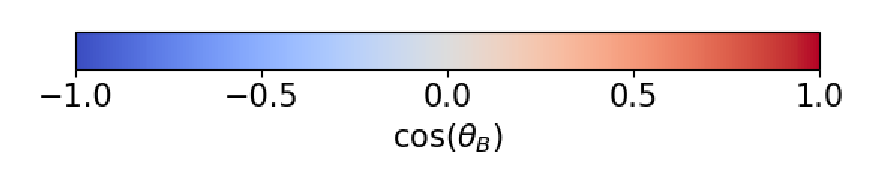}
            \includegraphics[width=0.49\hsize]{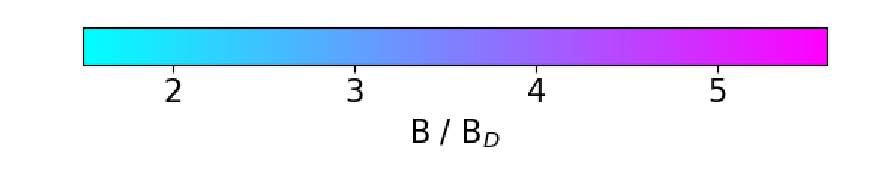}
        \end{subfigure}    
    \end{subfigure}
    \caption{Magnetic field configurations for a dipole with $(\alpha_D, \varphi_D) = (45^{\circ}, 0^{\circ})$ and a quadrupole with two different orientations. \emph{Left}: $(\alpha_Q, \varphi_Q) = (80^{\circ}, 95^{\circ})$. \emph{Right}: $(\alpha_Q, \varphi_Q) = (80^{\circ}, 180^{\circ})$. $B_Q/B_D = 5$ in both cases. \intrev{The top and middle panels show the angle of the surface magnetic field with respect to the radial direction, $\cos \theta_B$, and the relative magnetic field strength at the surface, $B/B_D$, in Mollweide projection. The maps cover the whole NS surface with latitudes in the range [$-90^\circ$, $90^\circ$] and longitudes in the range [$-180^\circ$, $180^\circ$].} The boundaries of the polar caps are shown with solid black lines. In the 3D representations, we represented closed field lines in blue, open field lines in red. The green arrow indicates the rotation axis of the NS. All distances are scaled by the stellar radius, $R^*$.}
    \label{fig:fieldlines}
\end{figure*}

To interpret the apparent field reconfiguration at the \gammacyg\ mode changes, we would need to explore the geometry of the magnetic field. \wtrev{In particular, we need to account for the small increase in the surface field $B_*$ implied from the pulsar fundamental plane, and for the phase change of the X-ray pulse observed by \citet{razzano23}.} For the purpose of this discussion, we limit ourselves to study an analytical model for the magnetic field in vacuum \citep{Kala2021}. The model assumes that the global magnetic field at large distances from the NS is a dipole but near the star a quadrupole field component dominates.  The dipole component of the field determines the position and geometry of the current sheet, where the $\gamma$-ray emission is produced.  This is consistent with the observed constancy of the $\gamma$-ray peak phases and only subtle changes in the profile. We set up a coordinate system with the origin in the NS center and the $z$ axis aligned to the rotation axis. We assume that the dipole (D) and quadrupole (Q) moments are both centered at the origin, and their orientations are each defined by a colatitude $\alpha$ and an azimuthal angle $\phi$. Without loss of generality, we always set the azimuthal angle of the dipole $\phi_D = 0$, and we \wtrevtwo{leave} the other angles free to vary. The ratio between the quadrupole and dipole surface fields is also free, and thus the full set of free parameters is $(\alpha_D, \alpha_Q, \phi_Q, B_Q/B_D)$. 
We simulate the quadrupole plus dipole configuration as follows. We start by creating a grid of $2^{16}$ equally spaced points on the NS surface. Each point corresponds to a unit surface area of \intrev{$2^{-16} \cdot 4 \pi R^{*2}$}, where $R^*$ is the NS radius. Starting from each point on the surface, we iteratively build a magnetic field line. At each iteration $i$ we compute the total magnetic field $\mathbf B_i$ at the point $\mathbf x_i$, then we make a step of size $0.01\ R^*$ in the direction of the field, \languagerev{that is,} $\mathbf x_{i+1} = \mathbf x_i\ +\ 0.01\ R^* \ \mathbf B_i / |\mathbf B_i|$. We stop when the path crosses the light cylinder (\emph{open} field lines) or goes back down and touches the surface (\emph{closed} field lines). For each line we record the strength and the inclination angle of the field at the surface. At the end of the simulation, we look for polar caps, defined as contiguous regions containing open field lines. For each polar cap we compute the area, the average magnetic field at the surface and the field inclination at the geometrical center.
We explore a range of orientations and relative strengths of the quadrupole, keeping the dipole constant.  We assume that the dipole has inclination of $\alpha_D = 45^\circ$ and our viewing angle is near $\zeta = 90^\circ$ \citep{pierbattista2015}.  We also explored the $\alpha_D = 60^\circ$ case and the results are very similar. \intrev{The choice of a constant $\alpha_D$ is supported by the fact that the peak positions are substantially stable, indicating that the dipole is fundamentally unaffected by the magnetospheric mode.} In general, the quadrupole has four poles, but in some configurations not all of them connect to the open field lines of the dipole. We assume that only polar caps that are open can generate current that heats the surface to produce thermal X-rays.  Searching for cases where the poles of the quadrupole that connect to the open dipole lines at larger altitude change by 0.21 in phase, as observed for the X-ray pulse, we focused on configurations where the quadrupole changes rotational phase  within a specific range but not inclination angle.  

Figure \ref{fig:fieldlines} shows two orientations of the quadrupole of relative strength $B_Q/B_D = 5$, \wtrev{$\phi_Q = 95^\circ$ and $\phi_Q = 180^\circ$}, that produce open polar caps seen by an observer at $\zeta \sim 90^\circ$ that are approximately 0.21 apart in phase.  In the Mollweide projection plots, the poles of the dipole are at $0^\circ$ and $180^\circ$ at $\zeta = 45^\circ$ latitude.  The first $\gamma$-ray peak, P1, would be at approximately 0.15 in phase or at $42^\circ$ (see \citet{Kala2014}).  The left-hand panels \intrev{represent} the field configuration and polar cap positions during interval B, where the X-ray pulse is at approximately 0.14 in phase later than the first $\gamma$-ray peak (see Figure 1 of \citet{razzano23}.  The right-hand panels represent the fields and polar caps for interval C, where the X-ray pulse is 0.35 in phase later than P1.  In the first case, three of the quadrupole poles connect to the dipole open field lines, and thus eventually to the current sheet, with two of the poles connected to the current sheet from one rotational hemisphere.  In the second configuration, only two poles of the quadrupole connect to the open dipole field.  The first case then would have three heated polar caps, only two of which are likely visible to one observer with $\zeta \sim 90^\circ$, and three polar caps supplying electron-positron pairs to the current sheet. 
The surface magnetic field strength is slightly lower over the poles in the first configuration (interval B) than in the second configuration (interval C), \intrev{while} the curvature of the field lines over the poles in interval B is smaller, since the two poles are connecting to the same pole of the dipole.  This could allow pairs to be created with smaller mean-free paths since the one-photon pair threshold depends on the angle of the photon to the magnetic field, as well as the field strength.  This configuration would therefore represent interval B, having a higher conductivity, a higher $\dot f$\languagerev{, and }a lower $\gamma$-ray flux.  
The P1 peak in the $\gamma$-ray profile comes from the part of the current sheet that is connected to two quadrupole poles in interval B and one quadrupole pole in interval C.  It would therefore be expected to show the largest change in flux and spectrum.  In fact we see in Figure \ref{fig:parameters} that P1 has the largest changes in flux and $E_p$ while P2 has much smaller, or no, changes in flux of $E_p$. 
\intrev{In both configurations, the estimated area of the polar cap near the equator is $\sim$$10^9$ cm$^2$, corresponding to a radius of $\sim$300 m. This is consistent with the size of the thermal emitting region observed by \citet{razzano23}.}

\languagerev{The cause of the mode changes is uncertain.}  \citet{takata20} suggest that strains on the stellar crust due to moving vortex lines in the superconducting core exceed the crustal elastic stresses, causing the crust to crack \citep{ruderman91}.  The crust will then move to relieve the stress, and they estimate that for an interval between successive displacements of $\tau \sim 7$ years for \gammacyg\, a crustal displacement of about \intrev{$1$ m} would result. \intrev{Our simple toy model outlined above assumes a quadrupole changing phase by $85^\circ$. This implies that a larger movement would be necessary to reconfigure the surface fields so that} the quadrupole poles disconnect or reconnect with the poles of the dipole.  However, our model is only illustrative and the real surface field could be more complex, containing higher multipoles allowing small displacements of the crust to cause large reconfiguration of the global field. \intrev{A long-term evolution of the magnetic field is also expected in highly magnetized NSs due to Lorentz forces acting on the electron superfluid component (\emph{Hall drift}, \citealt{GoldReis1992}). This effect favors the formation and evolution of small-scale structures in the internal current distribution, ultimately determining the geometry of the magnetic field. However, due to the complexity of Hall dynamics, we cannot yet say whether this effect could produce abrupt events such as the \gammacyg\ mode changes.}

Since \gammacyg\ is a radio-quiet pulsar, we should not see the radio pulse in either mode.  However, we must see one heated polar cap in each mode and it is believed that the radio emission is produced above the polar caps.  A possible reason for the non-detection of the radio pulses is that, if they are produced at altitudes of 50-100 stellar radii, the distortions of the field lines by the quadrupole cause the emission to be beamed in a direction away from the dipole \languagerev{and} quadrupole axes.  The X-ray emission is thermal, therefore more isotropic and is visible over a wider range of angles.  


\section{Conclusions}
\label{sec:5}

We have reported on the results of an updated \fermi\ analysis of the variable \gr\ pulsar \gammacyg. We performed a full phase-averaged and phase-resolved \gr\ spectral analysis, studying the properties of the \gr\ emission in detail. We investigated a \intrev{significant} mode change, which occurred around June 2020. This event produced an increase in the \gr\ flux by 16$\pm$2\% and a drop in the spin-down rate by 3.0$\pm$0.6\%. We also reported indications of \intrev{changes in the phase-averaged spectrum} and an increase in the pulsation significance concurrent with the flux drops. \intrev{The phase-resolved analysis suggests that the mode change affects the peaks of the light curve differently, implying an asymmetric variation in the structure of the magnetosphere.}

\intrev{In an attempt to interpret the nature of \gammacyg, we used the measured relative change in $\dot f$ to predict the observed flux drop. Under the hypothesis of pure CR emission in a dissipative magnetosphere, \journalrev{we computed the approximated flux variations produced by a change in the global conductivity}. The predictions are in disagreement with the observations, indicating that changes in the surface magnetic field also contribute to the flux modes. This hypothesis is consistent with the observed X-ray phase shift \citep{razzano23} observed across the 2014 mode change. For this reason, we invoked and explored a model for a multipolar magnetic field in vacuum, finding a configuration that qualitatively agrees with both the phase-averaged and the phase-resolved spectral analysis. However, this model does not account for the microphysics linking the fields to other electrodynamical quantities, and the derivation of a self-consistent radiative model to predict the \gammacyg\ spectral changes will be the topic of future work. Moreover, because the observed changes in the timing parameters differ from the classic glitch phenomenology, the mechanism driving the \gammacyg\ mode changes remains unknown. Therefore, although our interpretation is still qualitative, it should nevertheless pave the way to further research.}

In order to understand the physics of the mode changes more fully, it is essential to make observations of the X-ray pulse within each mode as well as between modes. \languagerev{NASA's Neutron Star Interior Composition Explorer (NICER; \citealt{nicer})} may be the ideal instrument to achieve an accurate absolute X-ray timing, \intrev{enabling joint X-ray and \gr\ light curve fitting.} Therefore, a continuous monitoring of \gammacyg\ \intrev{by \fermi} is the key to get further insight into the physics of variable \gr\ pulsars.


\begin{acknowledgements}
The \textit{Fermi} LAT Collaboration acknowledges generous ongoing support
from a number of agencies and institutes that have supported both the
development and the operation of the LAT as well as scientific data analysis.
These include the National Aeronautics and Space Administration and the
Department of Energy in the United States, the Commissariat \`a l'Energie Atomique
and the Centre National de la Recherche Scientifique / Institut National de Physique
Nucl\'eaire et de Physique des Particules in France, the Agenzia Spaziale Italiana
and the Istituto Nazionale di Fisica Nucleare in Italy, the Ministry of Education,
Culture, Sports, Science and Technology (MEXT), High Energy Accelerator Research
Organization (KEK) and Japan Aerospace Exploration Agency (JAXA) in Japan, and
the K.~A.~Wallenberg Foundation, the Swedish Research Council and the
Swedish National Space Board in Sweden.
 
Additional support for science analysis during the operations phase is gratefully
acknowledged from the Istituto Nazionale di Astrofisica in Italy and the Centre
National d'\'Etudes Spatiales in France. This work performed in part under DOE
Contract DE-AC02-76SF00515.

\wtrevtwo{Work at NRL is supported by NASA.}
\end{acknowledgements}


\bibliographystyle{aa}
\bibliography{mybib}
\end{document}